\documentclass[fleqn,usenatbib,referee]{mnras}
\usepackage{newtxtext,newtxmath}
\usepackage[T1]{fontenc}

\DeclareRobustCommand{\VAN}[3]{#2}
\let\VANthebibliography\thebibliography
\def\thebibliography{\DeclareRobustCommand{\VAN}[3]{##3}\VANthebibliography}

\usepackage{graphicx}	
\usepackage{amsmath}	
\usepackage{caption}
\usepackage{multirow}
\usepackage{comment}
\usepackage{booktabs}
\usepackage{color}
\renewcommand{\arraystretch}{1.1}

\title[High velocity stars in S-PLUS iDR4]{Characterisation of high velocity stars in the S-PLUS internal fourth data release}

\author[F. Quispe-Huaynasi et al.]{%
 F. Quispe-Huaynasi,$^{1}$\thanks{E-mail: fredifqh@gmail.com}
F. Roig,$^{1}$
V. M. Placco,$^{2}$
L. {Beraldo e Silva},$^{3}$
S. Daflon,$^{1}$
C. B. Pereira,$^{1}$
\newauthor
A. Kanaan,$^{4}$
C. Mendes de Oliveira,$^{5}$
T. Ribeiro,$^{6}$
W. Schoenell$^{7}$
\\
$^{1}$Observatório Nacional, MCTI, Rua Gal. José Cristino 77, Rio de Janeiro, 20921-400, RJ, Brazil \\
$^{2}$NSF's NOIRLab, 950 N. Cherry Ave., Tucson, AZ 85719, USA \\
$^{3}$Department of Astronomy, University of Michigan, 2074 East Hall, 530 Church St., Ann Arbor, MI 48109, USA \\
$^{4}$Departamento de Física, Universidade Federal de Santa Catarina, Florianópolis, 88040-900, SC, Brazil \\
$^{5}$Instituto de Astronomia, Geofísica e Ciências Atmosféricas, Universidade de São Paulo, Rua do Matão 1225, São Paulo, 05508-900, SP, Brazil \\
$^{6}$Rubin Observatory Project Office, 950 N. Cherry Ave., Tucson, AZ 85719, USA \\
$^{7}$GMTO Corporation, 465 N. Halstead Street, Suite 250, Pasadena, CA 91107, USA 
}

\date{Accepted 2023 November 20. Received 2023 November 15; in original form 2023 September 28}

\pubyear{2023}

\begin{document}
\label{firstpage}
\pagerange{\pageref{firstpage}--\pageref{lastpage}}
\maketitle

\begin{abstract}
In general, the atypical high velocity of some stars in the Galaxy can only be explained by invoking acceleration mechanisms related to extreme astrophysical events in the Milky Way. Using astrometric data from \textit{Gaia} and the photometric information in 12 filters of the S-PLUS, we performed a kinematic, dynamical, and chemical analysis of 64 stars with galactocentric velocities higher than 400~$\mathrm{km\,s}^{-1}$. All the stars are gravitationally bound to the Galaxy and exhibit halo kinematics. Some of the stars could be remnants of structures such as the Sequoia and the Gaia-Sausage/Enceladus. Supported by orbital and chemical analysis, we identified \textit{Gaia} DR3 5401875170994688896 as a star likely to be originated at the centre of the Galaxy. Application of a machine learning technique to the S-PLUS photometric data allows us to obtain very good estimates of magnesium abundances for this sample of high velocity stars.
\end{abstract}

\begin{keywords}
stars: fundamental parameters -- stars: kinematics and dynamics -- Galaxy: stellar content
\end{keywords}


\section{Introduction}

After the second data release of the \textit{Gaia} mission (\textit{Gaia} DR2), \citet{Marchetti2019} reported stars with high velocities, including some with a high probability of escaping the Milky Way (MW). In parallel, \citet{Hattori2018A} reported 30 stars with high velocities $( > 480\,\mathrm{km\,s}^{-1})$, with some stars in common with the ones reported by \citet{Marchetti2019}. Both works were performed on the sub-sample of stars in \textit{Gaia} DR2 with complete astrometric information ($\sim$ 7 million stars). But unlike \citet{Marchetti2019}, who searched for stars based on their total galactic velocity, \citet{Hattori2018A} selected stars based on their high tangential velocity and on quality cuts on the parallax. Based on their position in the colour-magnitude diagram (CMD), most of the reported stars would be metal-poor and belong to the red giant branch. Further analysis using high-resolution spectra confirmed that some of the candidates in these studies have the chemical properties of halo stars, and therefore belong to the high-velocity tail of the velocity distribution \citep{Hawkins2018, Reggiani2022}. Subsequently, searches have also been performed on later data releases of the \textit{Gaia} and new candidates have been reported \citep{marchetti2021, marchetti_2023, 2023ApJ...944L..39L}. 

Also, in the light of \textit{Gaia} data, several studies of confirmed and candidate unbound stars were reviewed. On one hand, \citet{2018MNRAS.479.2789B} showed that, with the exception of the star LAMOST J115209.12+120258.0, the late-type unbound candidate stars reported in the literature can be ruled out as candidates. On the other hand, young massive B stars reported as unbound \citep{2015ARA&A..53...15B} were reaffirmed as unbound \citep{2018A&A...620A..48I, 2018ApJ...866...39B}.

Furthermore, the search for high-velocity stars extended to the subset of stars lacking complete phase-space data in \textit{Gaia}'s records.  In this case, the astrometric information is complemented with radial velocities from spectroscopic surveys \citep{2018ApJ...863...87D, 2018ApJ...869L..31D, Li2021, Huaynasi2022, 2023AJ....166...12L}. The benefit of having stars in common between \textit{Gaia} and spectroscopic surveys lies in the ability to employ chemical data to restrict their origin, using chemical tagging \citep{Freeman2002}.

As expected, interest in this population of stars has grown in recent years with the advent of \textit{Gaia}. But the interest is not new. For example, some of the earliest studies related to high-velocity stars include the work of \citet{1922BAN.....1..133O, 1924PNAS...10..256O, 1926Obs....49..302O, 1930BAN.....5..189O}. These stars were identified by their high radial velocity or high proper motion, and several mechanisms were proposed for their origin \citep{2009ARep...53..839T}. Among the main mechanisms, we may cite: the ejection of stars at high velocity due to the rupture of a binary system by tidal forces exerted by the black hole at the centre of the Galaxy \citep[Hills mechanism,][]{Hills1988}; Type II and Type Ia supernova explosion in binary systems \citep{Blaauw1961, 2015Sci...347.1126G}; and gravitational interaction in dense systems as young clusters \citep{Poveda1967}. Depending on whether the velocity of these stars surpasses or falls short of the escape velocity, they can be categorised as either bound to or unbound to the Galaxy. On the other hand, they can be classified depending on their ejection mechanism: early OB type stars with origin in the galactic disc, as a result of type II supernovae or through dynamical interactions, are called runaway stars; stars as a result of type Ia supernovae are called hyper-runaway stars; stars originating through the Hills mechanism are called hyper-velocity stars (HVS). Recently, \citet{Evans2022a, Evans2022b} model the ejection of stars from the Galactic Centre (GC), exploring the possibility of detecting HVS as a function of the assumed stellar population around the GC. Comparing their results with the occurrence of HVS in the \textit{Gaia} DR2 and DR3, allow these authors to put strong constraints to the ejection rate of the Hill's mechanism, as well as on the stellar initial mass function (IMF) in the GC.

To constrain the local origin of these stars, and to discriminate between possible ejection mechanisms, stellar parameters and chemical information from stellar spectra play a key role \citep{Hawkins2018, Reggiani2022}. But stellar spectra, in particular, are not available for all the stars. On the other hand, multi-band photometric surveys like SkyMapper \citep{2007PASA...24....1K}, Pristine \citep{2017MNRAS.471.2587S}, J-PLUS \citep{2019A&A...622A.176C},  or miniJPAS \citep{2021A&A...653A..31B}, among others, equipped with filters strategically positioned on distinctive absorption and emission lines, are being employed to acquire stellar parameters for a vast array of stars and to identify peculiar astronomical objects for subsequent spectroscopic investigations.

In this work, we use deep learning algorithms and exploit the photometric information in 12 filters from the S-PLUS \citep[Southern
Photometric Local Universe Survey,][]{2019MNRAS.489..241M}\footnote{\url{https://www.splus.iag.usp.br/}}, to estimate effective temperature $T_\mathrm{{eff}}$, surface gravity $\log g$, and metallicity [Fe/H], for 64 high-velocity stars, as well as magnesium abundance [Mg/Fe], for the stars with [Fe/H] > -1, with the aim of constraining their possible origins. 

The article is structured as follows: the data set considered for our analysis is described in Section \ref{sect:data}. Section \ref{sect:selection} describes the high-velocity stars selection process. The kinematic and dynamical analysis, based on the stars' orbital parameters, dynamical variables, and kinematic spaces, is described in Section \ref{sect:kinematics}. The determination of stellar parameters using deep learning algorithms is presented in Section \ref{sect:parameters}. Finally, in Section \ref{sect:conclusions}, we present the conclusions. 

\section{Data sources}\label{sect:data}

The selection and kinematic characterisation of the sample of high-velocity stars have been performed using only the information provided by the \textit{Gaia} astrometric mission \citep{2016A&A...595A...1G}. The sample has been selected from the sample of stars with full phase-space information in the third release of data \citep[\textit{Gaia} DR3,][]{2022arXiv220800211G}.

Effective temperature, surface gravity and metallicity for the high velocity stars were estimated using deep learning algorithms trained with sources in common between the S-PLUS and the LAMOST \citep[Large Sky Area Multi-Object Fiber
Spectroscopic Telescope,][]{Cui2012}, and also with sources in common between the S-PLUS and the APOGEE \citep[Apache Point Observatory Galactic Evolution Experiment Data Release,][]{Abdurrouf2022}\footnote{\url{https://www.sdss4.org/dr17/}}. Specifically, to train and test the machine learning algorithms, we use data from the APOGEE DR17, and the Stellar Parameter Catalog of A, F, G, and K Stars, determined from low resolution spectra ($R=1\,800$), provided in LAMOST DR8.

S-PLUS is a photometric survey that collects data using the 12 optical filters listed in Table \ref{tab:jplus_filters}. Central wavelengths are located in important regions of emission and absorption lines, that provide valuable information about galaxies and stellar populations. The survey is being conducted from the Cerro Tololo Inter-American Observatory, in Chile, using a 0.8 m robotic telescope (T80-South) equipped with a camera (T80Cam) of $9.2\mathrm{k} \times 9.2\mathrm{k}$ pixels, that provides a $2\deg^2$ field of view (FoV), with a pixel scale of 0.55~$\mathrm{arsec\,pix}^{-1}$. Observational strategy, image reduction, and main scientific goals are presented in \citet{2019MNRAS.489..241M}. The S-PLUS iDR4 (internal Data Release 4), used in this study, includes 1\,629 fields, covering about 3\,000 square degrees of the southern sky, reduced and calibrated in all the survey bands. 

\begin{table}
\centering
\caption{S-PLUS photometric bands}
\label{tab:jplus_filters}
\begin{tabular}{lclc}
\hline 
Filter & Central wavelength & $\Delta \lambda$ & Spectral features \\
       &   \AA  &  \AA  &                                      \\
\hline
u     &   3563  & 352  & Balmer-break region                       \\
J0378 &   3770  & 151  & [OII]                                     \\
J0395 &   3940  & 103  & Ca\,H + K                                 \\
J0410 &   4094  & 201  & $\mathrm{H}_{\delta}$                     \\
J0430 &   4292  & 201  & G band                                    \\
g     &   4751  & 1545 & SDSS-like g                               \\
J0515 &   5133  & 207  &  Mg \textit{b} triplet                    \\
r     &   6258  & 1465 &  SDSS-like r                              \\
J0660 &   6614  & 147  &  $\mathrm{H}_{\alpha}$                    \\
i     &   7690  & 1506 &  SDSS-like i                              \\
J0861 &   8611  & 408  &  Ca triplet                               \\
z     &   8831  & 1182 &  SDSS-like z                              \\
\hline
\end{tabular}
\end{table}

Along the text, we identify the stars using their \textit{Gaia} ID rather than their S-PLUS ID, because this facilitates the cross-match with other surveys.

\section{High Velocity Stars Selection}\label{sect:selection}

Most of the stars with high velocity in the past were not identified by total space velocity, but by their high radial velocity or high proper motions. This was mainly due to the lack of information on stellar distances. However, with the Gaia data releases providing parallaxes and proper motions for a large number of stars, the search for stars with high velocity using total velocity became possible. In this work, we select stars with galactocentric velocity $V_{\mathrm{GC}}$ greater than 400~$\mathrm{km\,s}^{-1}$. We take this limit to avoid the high velocity tail of the velocity distribution of disc stars. The distances used to compute the $V_{\mathrm{GC}}$ are the photo-geometric distances estimated by \citet{2021yCat.1352....0B}, using a Bayesian approach. Our choice of this photo-geometric distances instead of the geometric distances (also estimated in \citealp{2021yCat.1352....0B}), is because they are more precise, especially for distant sources. But even if we consider the $V_{\mathrm{GC}}$ calculated using the geometric distances, it would not change the conclusions of this work. On the other hand, because $V_{\mathrm{GC}}$ is very sensitive to stellar distances, we discuss in the Appendix~\ref{apend:distances} the effect of the photo-astrometric distances estimated by \citet{2022A&A...658A..91A} on the values of $V_{\mathrm{GC}}$ for our sample.

In order to select sources with a reliable astrometric and photometric parameters in \textit{Gaia}, we only consider stars with positive parallax $\varpi > 0$, and stars flagged with the following labels in \textit{Gaia} DR3:  $\mathtt{ruwe} < 1.4$ and $-3 < \mathtt{astrometric\_gof\_al} < 3$, to ensure good astrometric solutions and to avoid binary systems, respectively. We also consider sources with $\mathtt{fidelity\_v2} > 0.5$, classified as good astrometric solution according to \citet{2022MNRAS.510.2597R}. Additionally, using $\mathtt{norm\_dg} < -3$, defined in \citet{2022MNRAS.510.2597R}, we remove stars with potential colour contamination from nearby sources\footnote{Originally, this flag was included in the external table \texttt{gedr3spur.main}, hosted at the German Astrophysical Virtual Observatory (GAVO), but now it is also available in the table \texttt{external.gaiaedr3\_spurious} at \url{https://gea.esac.esa.int/archive/}.}.

The transformation from the International Celestial Reference System (ICRS) coordinates to the Galactocentric reference frame (GRF) was performed using the \texttt{Pyia} package \citep{adrian-price-whelan_2018}. The parameters of the transformation are the ones set by default in version 4.0 of the \texttt{Astropy} Galactocentric frame: Sun's distance to the GC 8.122~kpc \citep{2018A&A...615L..15G}, Sun's height over the Galactic mid-plane 20.8~pc \citep{2019MNRAS.482.1417B}, and Sun's Cartesian velocity around the GC $(12.9,\, 245.6,\, 7.78)~\mathrm{km\,s}^{-1}$ \citep{2018RNAAS...2..210D}. The uncertainty propagation of the transformation is performed using 1\,000 Monte Carlo (MC) realisations, selected from multivariate normal distributions, $\mathcal{N}(\boldsymbol{\theta}, \Sigma)$, where $\boldsymbol{\theta} = (\alpha, \delta, \varpi, \mu_{\alpha}, \mu_{\delta}, v_\mathrm{rad})$ are the observables and $\Sigma$ is the covariance matrix provided by \textit{Gaia}\footnote{The elements of this matrix are the product of the uncertainties of two observables and their corresponding correlation coefficient. The correlation coefficients between $v_\mathrm{rad}$ and the other observables is assumed to be null, because they were obtained with different instruments.}. Finally, we consider the median and the 16th and 84th percentiles over the distributions to get the final positions and velocities in the GRF. 

After this process, we cross-match that sample with the S-PLUS iDR4 catalogue, resulting in a set of 64 stars. From this sample, all the stars have $\mathtt{fidelity\_v2} \sim 1$, therefore with a good astrometric solution. On the other hand, as warned by \citet{2022arXiv220605989B}, sources with $\mathtt{grvs\_mag - phot\_g\_mean\_mag} < -3$ may present problems in radial velocity measurements due to nearby bright sources. We verify that all stars in our sample have these values $>-3$. 

\begin{figure}
        \centering
        \includegraphics[scale=.75]{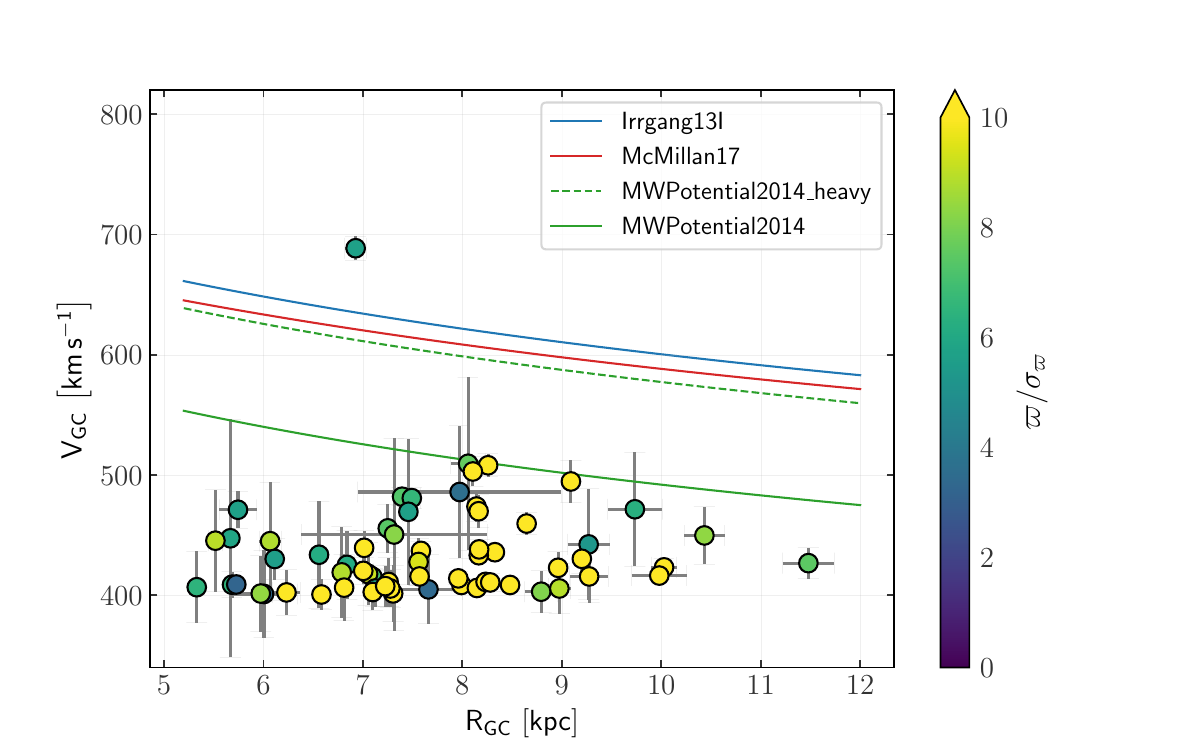}\caption{Galactocentric velocity as a function of Galactocentric distance, coded according to $\varpi/\sigma_{\varpi}$.}
        \label{fig:scape_velocity_curve}
\end{figure}

Figure \ref{fig:scape_velocity_curve} shows the Galactocentric velocity as a function of the Galactocentric radial distance $R_\mathrm{GC}$, in spherical coordinates. The solid lines in blue, red, and green in the figure represent the escape velocity curves calculated using the model I potential described in \citet{Irrgang2013}, the potential of \citet{McMillan2017}, and the MWPotential2014 potential of \citet{Bovy2015}, respectively. The green dashed line is the escape velocity curve assuming a higher mass of $\sim 1.2\times 10^{12}~M_{\odot}$ for the halo in the MWPotential2014 potential (MWPotential2014\_heavy), since its default value is underestimated compared to more recent estimates given in the literature \citep{2016ARA&A..54..529B, 2020SCPMA..6309801W}. The yellow dots are stars with $\varpi/\sigma_{\varpi} \geq 10$ (sources with good quality parallax measurements), the rest of the dots are stars with $\varpi/\sigma_{\varpi} < 10$ (sources with poor parallax). Except for the star \textit{Gaia} DR3 2690227738799604224, with $\varpi/\sigma_{\varpi} \sim 6$, all other stars are bound to the Galaxy according to the potentials considered (taking into account that the halo mass for the MWPotential2014 is underestimated). Looking at the astrometric parameters of the unbound candidate, we verify that the contribution to its high Galactocentric velocity comes from a radial velocity of 492~$\mathrm{km\,s}^{-1}$. As high radial velocities in \textit{Gaia} DR3 can represent spurious measurements due to very low signal to noise ratio ($\mathtt{rv\_expected\_sig\_to\_noise}$, SNR), as explained in \citet{2023A&A...674A...5K}, we verify that this star has $\mathtt{rv\_expected\_sig\_to\_noise} = 3.2 < 5$, and therefore there is a high chance of its radial velocity being spurious. It is worth mentioning that this is the most metal-poor star in the sample, thus the star with the weakest spectral lines.

\begin{table*}
\centering
\caption{Astrometric parameters from the \textit{Gaia} DR3. The last column reports the radial velocity from other spectroscopic surveys.}
\renewcommand*{\arraystretch}{1.05}
\label{tab:astrometric_parameters_splus}
\resizebox{\textwidth}{!}{%
\begin{tabular}{lrrcrrrrr}
\hline
\multicolumn{1}{c}{\textit{Gaia} DR3 ID}  &  \multicolumn{1}{c}{$\alpha$}  &  \multicolumn{1}{c}{$\delta$}  &  $\varpi$  &  \multicolumn{1}{c}{$\mu_{\alpha}$}  &  \multicolumn{1}{c}{$\mu_{\delta}$}  &  \multicolumn{1}{c}{$v_\mathrm{rad}$\,(Gaia)} & SNR &  \multicolumn{1}{c}{$v_\mathrm{rad}$\,(Spectroscopy)}                  \\
        &  \multicolumn{1}{c}{$\mathrm{(deg \pm mas)}$} & \multicolumn{1}{c}{$\mathrm{(deg \pm mas)}$} &  $\mathrm{(mas)}$  &  \multicolumn{1}{c}{$\mathrm{(mas\,yr^{-1})}$}  &  \multicolumn{1}{c}{$\mathrm{(mas\,yr^{-1})}$}  &  \multicolumn{1}{c}{$\mathrm{(km\,s^{-1})}$} &  &  \multicolumn{1}{c}{$\mathrm{(km\,s^{-1})}$} \\
\hline        
2320869248652782208  &    0.78  $\pm$  0.02  & -29.88  $\pm$  0.01  &  0.81  $\pm$  0.02  &   55.98  $\pm$  0.02  &   14.50  $\pm$  0.02  &  -220.80  $\pm$   5.04  & 10.8      &  -211.44  $\pm$  4.74 \, $^\mathrm{a\;\;\;}$   \\
2652780576075420544  &  340.85  $\pm$  0.01  &  -1.52  $\pm$  0.01  &  1.18  $\pm$  0.02  &   93.09  $\pm$  0.02  &  -86.11  $\pm$  0.01  &  -408.77  $\pm$   4.52  &  8.7      &                          \\
2688149803556948992  &  323.88  $\pm$  0.02  &   0.68  $\pm$  0.03  &  0.36  $\pm$  0.03  &  -25.77  $\pm$  0.03  &  -33.35  $\pm$  0.03  &  -122.77  $\pm$   5.03  &  5.7      &   125.65  $\pm$  3.48 \, $^\mathrm{b\;\;\;}$   \\
2688509099045891200  &  323.47  $\pm$  0.02  &   1.26  $\pm$  0.02  &  0.69  $\pm$  0.02  &   43.92  $\pm$  0.02  &   -8.82  $\pm$  0.02  &   166.37  $\pm$   8.46  &  4.9      &                          \\
2690227738799604224  &  316.49  $\pm$  0.02  &   0.49  $\pm$  0.02  &  0.14  $\pm$  0.03  &   -5.62  $\pm$  0.03  &   -9.30  $\pm$  0.02  &   491.77  $\pm$  10.01  &  3.2      &                          \\
3504215064633606912  &  196.14  $\pm$  0.02  & -22.94  $\pm$  0.01  &  0.18  $\pm$  0.02  &   12.42  $\pm$  0.02  &   -0.20  $\pm$  0.01  &     7.45  $\pm$   3.82  &  7.9      &                          \\
3504370095773053696  &  198.29  $\pm$  0.03  & -23.14  $\pm$  0.02  &  0.22  $\pm$  0.03  &    2.91  $\pm$  0.03  &    5.84  $\pm$  0.02  &   412.53  $\pm$   6.22  &  4.5      &                          \\
3525665333900712832  &  193.09  $\pm$  0.02  & -13.80  $\pm$  0.01  &  0.14  $\pm$  0.02  &  -17.69  $\pm$  0.02  &  -20.29  $\pm$  0.01  &   137.60  $\pm$   2.09  & 19.9      &                          \\
3545705651305437568  &  170.90  $\pm$  0.02  & -19.42  $\pm$  0.01  &  0.19  $\pm$  0.02  &   -4.70  $\pm$  0.02  &  -16.60  $\pm$  0.02  &   535.24  $\pm$   7.18  &  7.2      &                          \\
3546567427903076736  &  171.38  $\pm$  0.02  & -17.39  $\pm$  0.01  &  0.14  $\pm$  0.02  &   -8.17  $\pm$  0.02  &  -19.60  $\pm$  0.02  &   342.89  $\pm$   5.93  &  7.2      &                          \\
3612818501033473664  &  208.59  $\pm$  0.02  & -11.94  $\pm$  0.02  &  0.16  $\pm$  0.03  &    9.80  $\pm$  0.03  &   -5.31  $\pm$  0.02  &    46.24  $\pm$   8.36  &  4.6      &                          \\
3618819154461342592  &  209.20  $\pm$  0.03  &  -8.60  $\pm$  0.02  &  0.46  $\pm$  0.03  &    7.51  $\pm$  0.04  &  -54.41  $\pm$  0.02  &   158.63  $\pm$  10.77  &  2.7      &                          \\
3619582352969797888  &  208.35  $\pm$  0.01  &  -7.07  $\pm$  0.01  &  0.15  $\pm$  0.02  &  -19.83  $\pm$  0.02  &    0.38  $\pm$  0.02  &   195.24  $\pm$   0.73  & 17.0      &                          \\
3620532502815040768  &  208.98  $\pm$  0.02  &  -5.91  $\pm$  0.01  &  0.11  $\pm$  0.02  &    7.92  $\pm$  0.02  &   -5.90  $\pm$  0.02  &   -82.43  $\pm$   3.84  &  5.8      &                          \\
3630403952568941312  &  202.31  $\pm$  0.02  &  -7.86  $\pm$  0.01  &  0.17  $\pm$  0.02  &    7.90  $\pm$  0.02  &   -1.98  $\pm$  0.01  &   399.41  $\pm$   4.09  &  7.8      &   395.05  $\pm$  0.35 \, $^\mathrm{a\;\;\;}$   \\
3630462432843699200  &  202.48  $\pm$  0.01  &  -7.29  $\pm$  0.01  &  0.16  $\pm$  0.02  &    8.68  $\pm$  0.02  &  -12.50  $\pm$  0.01  &   161.15  $\pm$   1.07  & 22.4      &   162.94  $\pm$  0.19 \, $^\mathrm{a\;\;\;}$   \\
3630970956971494912  &  203.66  $\pm$  0.02  &  -6.57  $\pm$  0.01  &  0.14  $\pm$  0.02  &    7.07  $\pm$  0.02  &   -2.19  $\pm$  0.01  &   389.44  $\pm$   4.22  &  7.5      &                          \\
3631125674577631104  &  200.73  $\pm$  0.02  &  -7.70  $\pm$  0.02  &  0.15  $\pm$  0.03  &  -27.88  $\pm$  0.04  &   -5.80  $\pm$  0.02  &   -16.72  $\pm$   8.43  &  4.4      &                          \\
3657056526743053952  &  208.13  $\pm$  0.02  &  -3.79  $\pm$  0.01  &  0.38  $\pm$  0.02  &  -19.48  $\pm$  0.02  &  -48.43  $\pm$  0.02  &   257.82  $\pm$   5.20  &  6.2      &   255.64  $\pm$  3.42 \, $^\mathrm{b\;\;\;}$   \\
3696393857329932672  &  187.53  $\pm$  0.02  &  -0.46  $\pm$  0.01  &  0.12  $\pm$  0.02  &  -11.75  $\pm$  0.03  &  -19.72  $\pm$  0.02  &   156.04  $\pm$   7.82  &  4.5      &   146.37  $\pm$  2.60 \, $^\mathrm{a\;\;\;}$   \\
3750233780645824768  &  161.69  $\pm$  0.02  & -14.37  $\pm$  0.02  &  1.06  $\pm$  0.02  &  -67.03  $\pm$  0.03  &  -20.91  $\pm$  0.02  &  -273.42  $\pm$  10.79  &  3.4      &                          \\
3752558526183725440  &  155.65  $\pm$  0.01  & -14.06  $\pm$  0.01  &  1.12  $\pm$  0.01  &  -22.27  $\pm$  0.02  &  -85.83  $\pm$  0.01  &   568.97  $\pm$   1.24  & 20.9      &                          \\
3753526229559762176  &  154.09  $\pm$  0.02  & -13.28  $\pm$  0.02  &  0.75  $\pm$  0.02  &   -0.45  $\pm$  0.02  &   36.96  $\pm$  0.03  &   374.90  $\pm$   7.96  &  4.0      &   375.26  $\pm$  0.25 \, $^\mathrm{a\;\;\;}$   \\
3772348601612372992  &  153.14  $\pm$  0.02  &  -8.10  $\pm$  0.01  &  0.16  $\pm$  0.02  &   17.27  $\pm$  0.02  &   -3.64  $\pm$  0.02  &   143.23  $\pm$   3.76  &  7.6      &                          \\
3793377208170393984  &  173.49  $\pm$  0.02  &  -2.47  $\pm$  0.01  &  0.34  $\pm$  0.02  &   24.04  $\pm$  0.02  &   -8.42  $\pm$  0.01  &   296.89  $\pm$   2.33  & 12.9      &   291.52  $\pm$  3.13 \, $^\mathrm{b\;\;\;}$   \\
3891412241883772928  &  179.22  $\pm$  0.01  &   0.90  $\pm$  0.01  &  0.16  $\pm$  0.01  &   -8.73  $\pm$  0.02  &    9.82  $\pm$  0.01  &    59.98  $\pm$   1.44  & 18.6      &    63.06  $\pm$  4.38 \, $^\mathrm{b\;\;\;}$   \\
4227521019370523648  &  313.51  $\pm$  0.02  &   0.07  $\pm$  0.01  &  0.18  $\pm$  0.03  &   -6.84  $\pm$  0.03  &  -22.17  $\pm$  0.02  &  -471.28  $\pm$   4.90  &  6.4      &                          \\
4229238598267973376  &  305.50  $\pm$  0.01  &  -1.31  $\pm$  0.01  &  0.06  $\pm$  0.02  &    7.34  $\pm$  0.02  &    1.59  $\pm$  0.01  &  -126.33  $\pm$   0.71  & 34.2      &                          \\
4235951322918940928  &  300.18  $\pm$  0.01  &  -1.27  $\pm$  0.01  &  0.06  $\pm$  0.02  &   -4.54  $\pm$  0.02  &  -10.69  $\pm$  0.01  &  -442.24  $\pm$   1.00  & 18.1      &                          \\
4236552309105405824  &  301.29  $\pm$  0.02  &  -0.18  $\pm$  0.02  &  0.15  $\pm$  0.03  &   -7.24  $\pm$  0.03  &  -20.72  $\pm$  0.02  &   -87.84  $\pm$   5.97  &  4.8      &                          \\
4242563614044957696  &  304.34  $\pm$  0.02  &   1.04  $\pm$  0.02  &  0.56  $\pm$  0.02  &   -0.10  $\pm$  0.02  &   -6.77  $\pm$  0.02  &   213.79  $\pm$  17.44  &  4.4      &                          \\
4681680795508995840  &   64.11  $\pm$  0.01  & -58.86  $\pm$  0.01  &  1.65  $\pm$  0.01  &  156.31  $\pm$  0.02  &   14.81  $\pm$  0.02  &   461.89  $\pm$   5.05  &  8.5      &                          \\
4787052870314603520  &   72.41  $\pm$  0.01  & -47.05  $\pm$  0.02  &  0.38  $\pm$  0.02  &   23.38  $\pm$  0.02  &  -32.20  $\pm$  0.02  &   -83.90  $\pm$   8.52  &  4.0      &                          \\
4788482544668341632  &   66.47  $\pm$  0.02  & -49.11  $\pm$  0.02  &  0.12  $\pm$  0.02  &    7.31  $\pm$  0.02  &   -8.28  $\pm$  0.02  &  -186.67  $\pm$   9.71  &  4.0      &                          \\
4788911315546547328  &   66.00  $\pm$  0.01  & -47.62  $\pm$  0.01  &  0.20  $\pm$  0.01  &    9.88  $\pm$  0.01  &  -11.99  $\pm$  0.02  &  -200.44  $\pm$   4.15  &  8.0      &                          \\
4790354875531489024  &   71.09  $\pm$  0.01  & -45.23  $\pm$  0.01  &  2.43  $\pm$  0.01  &   61.72  $\pm$  0.01  &  139.09  $\pm$  0.02  &   -45.36  $\pm$   3.16  & 19.3      &                          \\
4856695936817617280  &   56.14  $\pm$  0.01  & -37.53  $\pm$  0.01  &  0.39  $\pm$  0.01  &   59.00  $\pm$  0.01  &  -16.68  $\pm$  0.01  &    44.27  $\pm$   2.54  & 10.0      &                          \\
4882090566610735360  &   63.19  $\pm$  0.01  & -34.22  $\pm$  0.01  &  0.42  $\pm$  0.01  &   56.00  $\pm$  0.01  &  -15.59  $\pm$  0.01  &   216.37  $\pm$   2.32  & 14.8      &                          \\
5048231102894377344  &   48.73  $\pm$  0.01  & -33.85  $\pm$  0.01  &  0.25  $\pm$  0.02  &    2.81  $\pm$  0.01  &  -11.48  $\pm$  0.02  &  -280.14  $\pm$   3.43  &  9.0      &                          \\
5401875170994688896  &  164.73  $\pm$  0.01  & -34.30  $\pm$  0.01  &  0.15  $\pm$  0.02  &  -18.43  $\pm$  0.01  &    3.45  $\pm$  0.02  &   364.05  $\pm$   4.34  &  8.3      &                          \\
5441944638890204160  &  156.68  $\pm$  0.01  & -37.87  $\pm$  0.02  &  1.59  $\pm$  0.02  &   -9.50  $\pm$  0.02  &   15.99  $\pm$  0.02  &  -260.68  $\pm$  13.10  &  2.5      &                          \\
5463127795547217024  &  149.84  $\pm$  0.01  & -31.00  $\pm$  0.02  &  2.52  $\pm$  0.02  & -148.59  $\pm$  0.02  &  150.91  $\pm$  0.02  &   128.18  $\pm$   5.80  &  3.4      &                          \\
5469865053045546112  &  160.51  $\pm$  0.01  & -25.78  $\pm$  0.02  &  0.60  $\pm$  0.02  &  -35.40  $\pm$  0.02  &   28.15  $\pm$  0.02  &   409.73  $\pm$   2.82  &  7.5      &                          \\
5473218563510860032  &  153.57  $\pm$  0.03  & -24.71  $\pm$  0.03  &  5.72  $\pm$  0.04  &  -39.00  $\pm$  0.05  &   -8.13  $\pm$  0.06  &  -246.43  $\pm$  13.35  &  3.2      &                          \\
5475847186574041984  &  157.50  $\pm$  0.02  & -21.53  $\pm$  0.02  &  2.56  $\pm$  0.03  &  -52.45  $\pm$  0.03  &    6.74  $\pm$  0.03  &  -244.74  $\pm$  14.63  &  2.7      &                          \\
5674511482965847168  &  152.17  $\pm$  0.03  & -14.83  $\pm$  0.03  &  4.96  $\pm$  0.03  &  -20.07  $\pm$  0.03  &   10.95  $\pm$  0.03  &  -213.34  $\pm$   8.79  &  2.9      &                          \\
6270952693712299008  &  214.43  $\pm$  0.02  & -25.74  $\pm$  0.02  &  0.25  $\pm$  0.02  &   12.67  $\pm$  0.03  &  -14.31  $\pm$  0.03  &   229.77  $\pm$   8.03  &  4.9      &                          \\
6478130959611566720  &  315.61  $\pm$  0.01  & -49.36  $\pm$  0.01  &  0.09  $\pm$  0.02  &    0.80  $\pm$  0.02  &  -11.66  $\pm$  0.01  &  -404.70  $\pm$   3.72  &  9.1      &  -386.95  $\pm$  0.94 \, $^\mathrm{a\;\;\;}$   \\
6478219813895286912  &  313.91  $\pm$  0.02  & -49.91  $\pm$  0.01  &  0.18  $\pm$  0.02  &  -15.88  $\pm$  0.02  &  -21.13  $\pm$  0.02  &   -66.35  $\pm$   5.18  &  6.3      &                          \\
6503261122497594624  &  341.80  $\pm$  0.01  & -59.56  $\pm$  0.01  &  0.14  $\pm$  0.01  &   14.18  $\pm$  0.01  &    0.69  $\pm$  0.01  &  -107.47  $\pm$   1.23  & 20.3      &                          \\
6575967187877774720  &  323.25  $\pm$  0.01  & -45.48  $\pm$  0.01  &  0.12  $\pm$  0.01  &   -6.51  $\pm$  0.01  &   -9.77  $\pm$  0.01  &  -209.43  $\pm$   0.54  & 27.8      &  -210.15  $\pm$  0.12 \, $^\mathrm{a\;\;\;}$   \\
6580745253096436224  &  319.26  $\pm$  0.01  & -41.68  $\pm$  0.01  &  0.13  $\pm$  0.02  &    3.43  $\pm$  0.02  &    6.04  $\pm$  0.02  &   -59.33  $\pm$   5.96  &  8.1      &                          \\
6617426988221664640  &  327.26  $\pm$  0.02  & -29.32  $\pm$  0.01  &  0.06  $\pm$  0.02  &   -7.37  $\pm$  0.02  &   -8.26  $\pm$  0.02  &  -197.70  $\pm$   3.44  &  7.8      &                          \\
6619340275892884224  &  331.23  $\pm$  0.01  & -27.34  $\pm$  0.01  &  0.15  $\pm$  0.02  &   17.61  $\pm$  0.02  &  -17.80  $\pm$  0.02  &  -136.05  $\pm$   1.02  & 20.0      &                          \\
6625197335678814208  &  334.07  $\pm$  0.01  & -25.56  $\pm$  0.01  &  0.23  $\pm$  0.02  &   -6.98  $\pm$  0.02  &  -27.35  $\pm$  0.01  &  -387.28  $\pm$   2.50  & 13.1      &                          \\
6775627229489106688  &  314.51  $\pm$  0.01  & -36.26  $\pm$  0.01  &  0.13  $\pm$  0.02  &  -14.45  $\pm$  0.02  &   -9.21  $\pm$  0.02  &    87.08  $\pm$   4.37  &  6.4      &                          \\
6776360157068155520  &  319.07  $\pm$  0.02  & -35.91  $\pm$  0.01  &  0.09  $\pm$  0.03  &   -3.33  $\pm$  0.03  &   -1.38  $\pm$  0.02  &  -363.68  $\pm$   5.11  &  6.1      &                          \\
6782909879114518016  &  320.93  $\pm$  0.02  & -33.33  $\pm$  0.02  &  0.22  $\pm$  0.03  &   11.30  $\pm$  0.02  &  -25.29  $\pm$  0.02  &  -348.97  $\pm$   4.44  &  6.1      &                          \\
6785190193214463872  &  322.96  $\pm$  0.03  & -30.63  $\pm$  0.02  &  0.17  $\pm$  0.03  &    8.23  $\pm$  0.03  &    4.92  $\pm$  0.02  &     0.14  $\pm$   5.97  &  4.4      &                          \\
6785833613675654912  &  322.05  $\pm$  0.02  & -30.72  $\pm$  0.01  &  0.64  $\pm$  0.02  &  -23.55  $\pm$  0.02  &  -79.01  $\pm$  0.02  &  -243.08  $\pm$   4.45  &  9.9      &                          \\
6790317563828425728  &  319.28  $\pm$  0.02  & -28.24  $\pm$  0.01  &  0.19  $\pm$  0.02  &  -16.29  $\pm$  0.02  &  -18.45  $\pm$  0.02  &    42.07  $\pm$   4.41  &  9.2      &                          \\
6790795542148925952  &  318.90  $\pm$  0.02  & -26.88  $\pm$  0.01  &  0.33  $\pm$  0.02  &   28.62  $\pm$  0.02  &   -3.94  $\pm$  0.01  &    41.72  $\pm$   3.30  & 13.1      &                          \\
6812361977394894464  &  330.09  $\pm$  0.01  & -25.04  $\pm$  0.01  &  0.25  $\pm$  0.01  &   -0.39  $\pm$  0.01  &  -37.02  $\pm$  0.01  &   -26.46  $\pm$   0.80  & 23.6      &                          \\
6917206978109246080  &  315.39  $\pm$  0.02  &  -1.05  $\pm$  0.02  &  0.94  $\pm$  0.03  &   -7.29  $\pm$  0.03  &   -3.26  $\pm$  0.02  &   214.45  $\pm$   8.72  &  3.8      &                          \\
\hline 
\multicolumn{9}{l}{$^\mathrm{a}$ GALAH DR3 catalogue}\\
\multicolumn{9}{l}{$^\mathrm{b}$ LAMOST DR8 LR catalogue}\\
\end{tabular}
}
\end{table*}

To check if the stars in the sample have radial velocity information in other catalogues, we performed a cross-match with spectroscopic catalogues. Unfortunately, the unbound candidate star has no spectrum available in archives of observed data, and there is currently no way to obtain a better estimate of its radial velocity. But we found radial velocity information for 7 stars in GALAH DR3, and for 4 stars in the LAMOST DR8 LR catalogue. Astrometric parameters from \textit{Gaia} and radial velocities from spectroscopic surveys are reported in Table \ref{tab:astrometric_parameters_splus}. Comparing the radial velocities from spectroscopic surveys with the ones provided by \textit{Gaia} DR3, we find that the mean difference is $\sim -1\,\mathrm{km\,s}^{-1}$, with a dispersion of $7\,~\mathrm{km\,s}^{-1}$, showing a good agreement between the different catalogues. 

Table \ref{tab:astrometric_parameters_splus} shows that the spectra of some targets have very low SNR, and therefore no reliable radial velocity in \textit{Gaia}. In this case, the typical approach involves a cutoff in the SNR, but the threshold in SNR is not well defined. For instance, if we set a $\mathrm{SNR} > 5$, following \citet{2022MNRAS.515..767M}, we would exclude various stars, including \textit{Gaia} DR3 3753526229559762176 and \textit{Gaia} DR3 3696393857329932672, whose radial velocities in Gaia are consistent with radial velocities from spectroscopic surveys. Therefore, we will not adopt any cut in SNR, but conclusions about these stars will be discussed in the light of their stellar parameters in Section \ref{sect:chemical}. 
After these considerations, our final sample consists of 64 stars, and hereafter we will refer to it as the HiVel sample.

\section{Kinematics and dynamics}\label{sect:kinematics}

Orbital parameters, kinematic variables and dynamical variables are fundamental to understand the kinematic nature of stars. This section describes how the orbital parameters are calculated and then presents three of the most important kinematic spaces for the HiVel sample.

\subsection{Orbit integration}\label{sect:orbit_integration}

With position and velocity information, calculated from the astrometric parameters given by \textit{Gaia}, and considering a gravitational potential for the Galaxy, we calculate the orbital parameters applying backwards orbital integration over 10~Gyr in the past, using the \texttt{Galpy} package \citep{Bovy2015}\footnote{\url{https://docs.galpy.org/en/v1.9.0/}}. The orbit integration is performed for each of the MC realisations of each star (described in the previous section), and the final values of the parameters are given by the median, the 16th and 84th percentiles. We adopted the gravitational potential of \citet{McMillan2017}. This model is composed of a bulge, a dark matter halo, a thin disc, a thick disc, and an HI and molecular gas disc. All the components are axisymmetric and are implemented in \texttt{Galpy}. 

\begin{figure}
        \centering
        \includegraphics[width=\columnwidth]{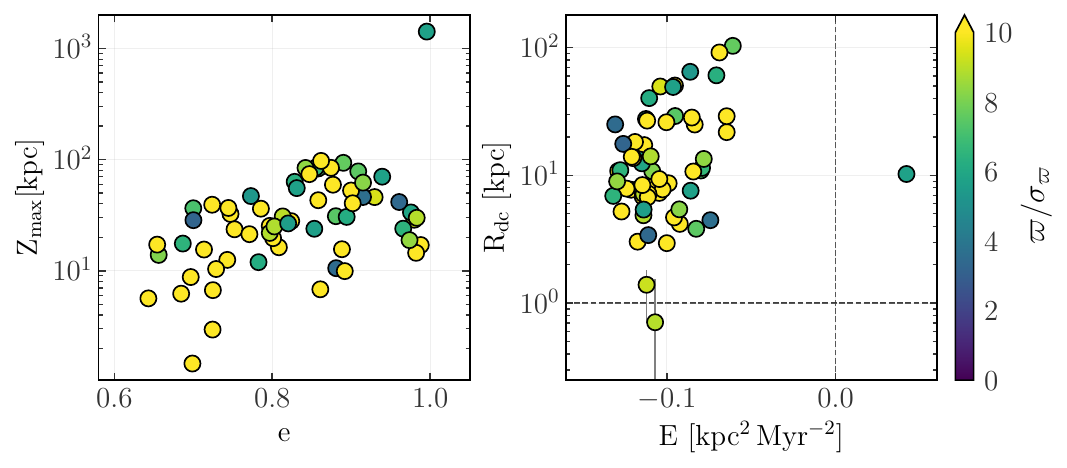}
        \vspace{-0.5cm}
        \caption{\textit{Left:} Maximum height above the Galactic plane as a function of eccentricity. \textit{Right:} Galactocentric distance to the point of the last crossing through the disc ($\mathrm{R_{dc}}$, $Z=0$) as a function of orbital energy. The vertical line shows the limit between bound and unbound stars, and the horizontal line corresponds to 1~kpc. The colour code is the same as in Fig.~\ref{fig:scape_velocity_curve}.}
        \label{fig:orb_param}
\end{figure}

The left panel of Fig.~\ref{fig:orb_param} shows the maximum height relative to the galactic plane $Z_\mathrm{{max}}$, as a function of eccentricity $e$. It is clear from the figure that the orbits are eccentric $(e > 0.5)$ and that most stars have $Z_\mathrm{{max}} > 5$~kpc. To evaluate a possible origin at the centre of the Galaxy, the right panel of the Fig.~\ref{fig:orb_param} plots the relation between the orbital energy $E$ and the distance from the GC to the point of last disc crossing $R_\mathrm{{dc}}$ (i.e. the galactocentric distance when $Z=0$ for the last time).
We see that the stars \textit{Gaia} DR3 3619582352969797888, with $R_\mathrm{{dc}} < 1$~kpc, and  Gaia DR3 5401875170994688896, with $R_\mathrm{{dc}} =  1.4$~kpc, have the highest probability among the sample of having originated in the GC.
The vertical dashed line represents the boundary between the bound $(E < 0)$ and unbound $(E > 0)$ stars. 

We also investigate the possible influence of the Galactic bar in the orbital properties derived for our sample, which might be specially important for orbits crossing the disc close to the centre. We integrate orbits for the same stars, but in the barred potential of \citet[also Hunter et al. in prep.]{2022MNRAS.514L...1S}. This potential is a good fit to both the inner and outer parts of the MW potential and it is available in the \texttt{AGAMA} \citep[Action-based Galaxy Modelling Architecture,][]{2019MNRAS.482.1525V} repository\footnote{\url{https://github.com/GalacticDynamics-Oxford/Agama}}. We assume a pattern speed $\Omega_P =35.5\,\mathrm{km\,s^{-1}\,kpc^{-1}}$, a bar inclination angle of $30^{\circ}$, and again we integrate for $-10$~Gyr. From the analysis of the orbital trajectories in this gravitational potential, we see that the distance from the GC to the point of the last disc crossing, for the two stars that might be originated in the GC, is $\sim1$~kpc, as in the previous simulations without a bar. In Section \ref{sect:chemical}, we will re-discuss the possible origin of these two stars taking into account their position in the [Mg/Fe] vs. [Fe/H] diagram.

\subsubsection{LMC origin}

Since the ejection mechanisms mentioned before can also operate in the Large Magellanic Cloud (LMC), stars with high velocities may have formed in this system \citep{2008A&A...480L..37P, 2017A&A...603A..75L, 2019MNRAS.483.2007E}. To determine if any of the stars in the HiVel sample might have come from the LMC, we calculate the orbit of the LMC as a test particle under dynamical friction. To compute the effect of dynamical friction, we consider the mass of $1.5\times 10^{11}M_\odot$ for the LMC, as adopted in \citet{2019MNRAS.483.2007E}, together with the McMillan17 potential, and the half-mass radius (rhm) of 6~kpc in the dynamical friction formula implemented in \texttt{Galpy} (ChandrasekharDynamicalFrictionForce). The rhm value was selected in order to match the orbit of the LMC to the orbit given by \citet{2023Galax..11...59V}. However, as detailed in that paper, the trajectory of the LMC is very sensitive to the mass of the MW, the mass of the LMC, the shape of the halo, and uncertainties in the present-day position and velocity measurements. In this work, the phase-space information for the LMC was calculated using astrometric data from \citet{2021A&A...649A...7G} and the distance given by \citet{2019Natur.567..200P}. To determine the probability of the origin of a star in the LMC, we calculate the mutual distance $d_\mathrm{{star - LMC}}$ between the orbit of the star and the orbit of the LMC at each integration step, and check if they have a close encounter, $d_\mathrm{{star - LMC}} < 5$~kpc, in the past. This is done for the 1\,000 MC realisations of each star. We find that $d_\mathrm{{star - LMC}} > 5$~kpc in all cases, and conclude that all the stars have a very low probability of having been ejected from the LMC.

\subsection{Kinematical spaces}

Three of the most commonly used kinematic spaces in the literature to understand the kinematic nature of stars are described below for the HiVel sample. Integrals of motion used to construct these kinematic planes were calculated with \texttt{Galpy}, using the McMillan's gravitational potential.

\begin{figure}
        \centering
        \includegraphics[width=\columnwidth]{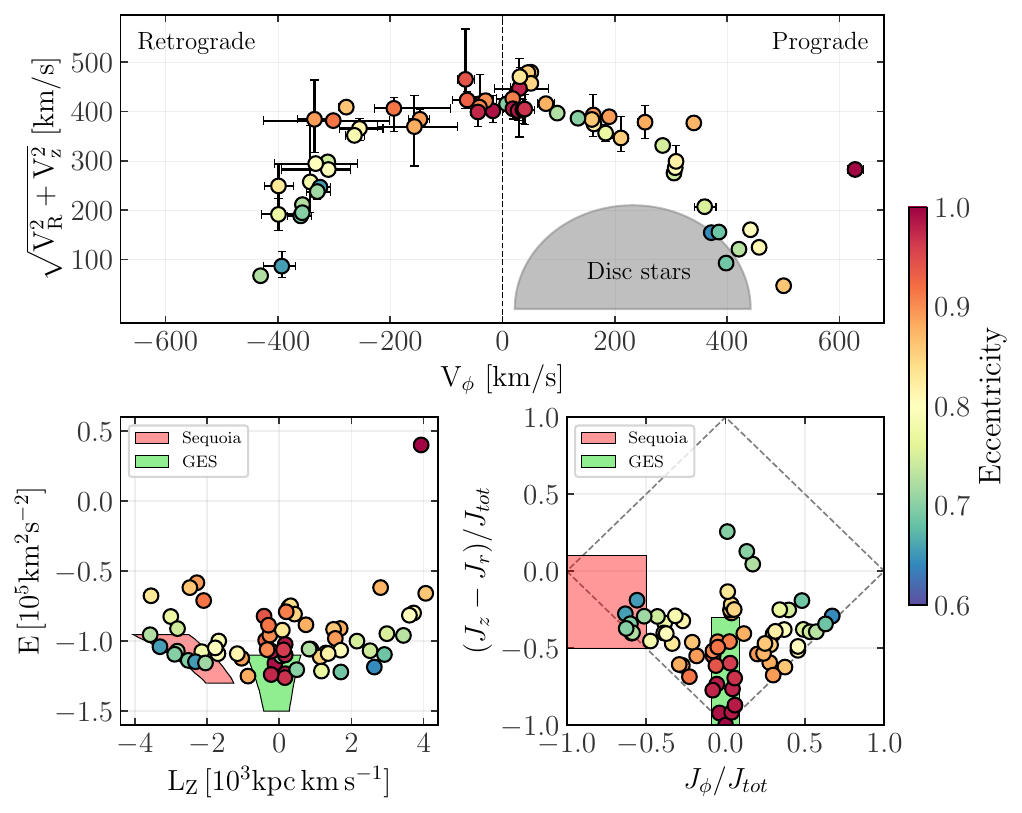}
        \vspace{-0.5cm}
        \caption{Toomre diagram (top), Lindblad diagram (left bottom), and scaled actions diamond diagram (right bottom). The stars are colour-coded according to eccentricity. The shaded region in the Toomre diagram is the approximate locus of the disc stars. The areas in green and magenta in the bottom panels correspond to the approximate locations of the Gaia-Enceladus/Sausage and the Sequoia mergers remnants, respectively.}
        \label{fig:kinematical_space}
\end{figure}

The first kinematic space is the Toomre diagram (upper panel of Fig.~\ref{fig:kinematical_space}). Considering that the halo and disc stellar population possess different kinematic properties, this diagram is widely used in the literature to kinematically classify these two populations \citep{2003A&A...410..527B, Nissen2010, 2015MNRAS.447.2046H}. Unlike the kinematic spaces described below, which use integrals of motion calculated by assuming a gravitational potential for the MW, the Toomre space is built purely from velocity information, considering the galactocentric velocities in cylindrical coordinates. The azimuthal galactocentric velocity ($V_{\phi}$) is plotted on the x-axis and the quantity $\sqrt{V_{R}^2 + V_{Z}^2}$ is plotted on the y-axis. Because the stars in the disc have nearly circular orbits and low velocity dispersion, they lie in the shaded region defined for stars with $|V_{\mathrm{GC}}-V_{\mathrm{LSR}}| < 210\,\mathrm{km\,s}^{-1}$, with $V_{\mathrm{LSR}} = 232\,\mathrm{km\,s}^{-1}$ being the velocity of the Local Standard of Rest (LSR), adopted from \citet{2018ApJ...860L..11K}. On the other hand, with a larger contribution of the vertical and radial velocities, the halo stars will be outside this region, $|V_{\mathrm{GC}}-V_{\mathrm{LSR}}| > 210\,\mathrm{km\,s}^{-1}$. It is worth noting, however, that the boundary between these two populations is not well defined. With the exception of a few stars, which lie close to the separation boundary between these two populations, the majority of our HiVel stars clearly exhibit the kinematic behaviour of the halo stars. This is further supported by the fact that the distribution of $V_\phi$ has approximately equal numbers of prograde and retrograde stars.

The second kinematical space (left bottom panel of Fig.~\ref{fig:kinematical_space}) is the energy vs. angular momentum plane, also known as Lindblad diagram. The two quantities $E,\,L_Z$ used in this diagram, are constants in axisymmetric potentials. This space concentrates disc and halo stars into specific regions. Unlike halo stars, which are mainly located in regions of low angular momentum and high orbital energy, disc stars will occupy regions of positive angular momentum ($L_Z > 0$) and moderate orbital energy \citep{2022MNRAS.510.5119L}. However, as in Toomre's diagram, there is a region of overlap between these two populations. With complete phase space data from a substantial sample of stars, this space was also used to identify stellar structures in the Galactic halo \citep{2020ARA&A..58..205H}. However, the regions occupied by these structures in the Lindblad diagram are not well defined, and some of them may overlap \citep[see][]{2020ApJ...901...48N, 2021ApJ...909L..26B}. We can see that some HiVel stars are in regions expected for structures such as Sequoia (magenta shaded area), that has stars with retrograde motion, and the Gaia-Enceladus/Sausage (GES) structure (green shaded region), which is composed of stars with prograde and retrograde motion. The borders of the shaded regions were extracted from \citet{2019A&A...631L...9K}.
As expected from the previous section, star \textit{Gaia} DR3 2690227738799604224 has positive orbital energy ($E > 0$), indicating an unbound candidate. 

The third kinematical space is the so-called action diamond (right-bottom panel of Fig.~\ref{fig:kinematical_space}). The actions $J_R, J_{\phi}, J_Z$ are integrals of motion of the system. In axisymmetric potentials, in particular, $J_{\phi} = L_Z$, but $J_R$ and $J_Z$  need to be estimated numerically. The actions for our HiVel stars have been calculated using the Stäckel approximation \citep{2012MNRAS.426.1324B}. This diagram allows to segregate stellar populations \citep{2022MNRAS.510.5119L}, and to identify substructures in the Galactic halo \citep{2019MNRAS.488.1235M, 2020ApJ...901...48N}. As shown in Fig.~\ref{fig:kinematical_space}, stars with eccentric orbits tend to group in the lower corner of the diamond, while stars with retrograde and prograde motions concentrate in the left and right corners, respectively. The region in magenta in this diagram corresponds to the location of the remnants of the Sequoia, with highly retrograde stars, and the region in green corresponds to the location of the remnants of the GES merger, with stars in highly eccentric orbits \citep{2019MNRAS.488.1235M}. The limits of these regions are not well defined and we adopt the ones given in \citet{2020MNRAS.497.1236M}. We can see that some stars of the HiVel sample lie within or very close to these two regions. In principle, a detailed analysis of the chemical abundances may help to confirm whether these stars belong to these populations, although this is still an open problem \citep{2019A&A...631L...9K, 2021ApJ...908L...8A, 2023MNRAS.520.5671H}

\section{Stellar parameters}\label{sect:parameters}

\subsection{Parameters determined by Convolutional Neural Network}\label{sect:CNN_pars}

With the development and implementation of machine learning algorithms, and access to large amounts of data from space telescopes and ground-based telescopes in astronomy, the application in the different areas of astronomy was expected  \citep[see][and references therein]{2022arXiv221201493D, 2023RSOS...1021454S}. Here we describe the process to obtain stellar parameters for the HiVel sample using a Convolutional Neural Network (CNN) architecture, the photometric information in 12 filters from S-PLUS, and the stellar parameters from LAMOST. CNN is a deep learning algorithm frequently used in computer vision. However, CNNs can also be used for one-dimensional sequential data \citep{2021MSSP..15107398K}. As an application in astronomy, \citet[and references therein]{2022OAst...31...38G} shows how CNN architectures can be used for the determination of stellar parameters through stellar spectra.

Since the order of the features is important in the CNN architecture for pattern recognition, we order the magnitudes according to increasing wavelength. The labels of the data (output parameters) are the values of $T_\mathrm{eff}$, [Fe/H], and $\log\,g$.

\subsection{Data training and testing}

The data set used to train and test the CNN comes from the data in common between S-PLUS iDR4 and LAMOST DR8 ($\sim 60\,000$ stars). Before this process we perform the following cuts:
\begin{enumerate}
        \item For the S-PLUS data, we consider stars with apparent magnitudes $\mathtt{mag\_auto} < 20$ and associated uncertainties $\mathtt{mag\_err\_auto} < 0.15$, in all the photometric bands. This is a compromise between the quality and the quantity of the data for training and testing the CNN. In particular, it guarantees enough training/testing data in the $\mathrm{[Fe/H]}<-1$ region ($\sim 3\,500$ stars). Imposing a more strict condition, like $\mathtt{mag\_err\_auto} < 0.05$, reduces the data sample in the $\mathrm{[Fe/H]}<-1$ region by about 20\%.
        \item For the LAMOST data, we select stars with temperatures in the range $3\,500 \leq T_\mathrm{eff} \leq 8\,000$~K, with uncertainties of $< 150$~K. In addition, we consider sources with uncertainties $<0.2$~dex both in $\log g$ and [Fe/H].
\end{enumerate}
By applying these quality cuts, we obtain a sample consisting of about 55 000 stars, of which 80\% is used to train and 20\% to test the CNN.

The initial CNN architecture for training the algorithm is described in \citet{Quispe2023}. We use three one-dimensional convolutional layers to build the features map from the input data. A one-dimensional maxpooling layer is associated to each of the convolutional layers, in order to reduce the number of inputs that will feed the neural network. The network itself is composed of an input layer, three hidden layers, and an output layer. The Relu activation function is used in both the convolutional layers and the network layers. We select the Adam optimisation algorithm \citep{2014arXiv1412.6980K}, and the mean absolute error (mae) as loss function. 

The final hyperparameters of the CNN, learning rate, number of kernels, kernel size, and number of neurons in each hidden layer are obtained via the \texttt{keras} tuner \citep{omalley2019kerastuner}, using the Bayesian search optimiser.

\subsubsection{Model testing}

\begin{figure*}
        \centering
        \includegraphics[width=\textwidth]{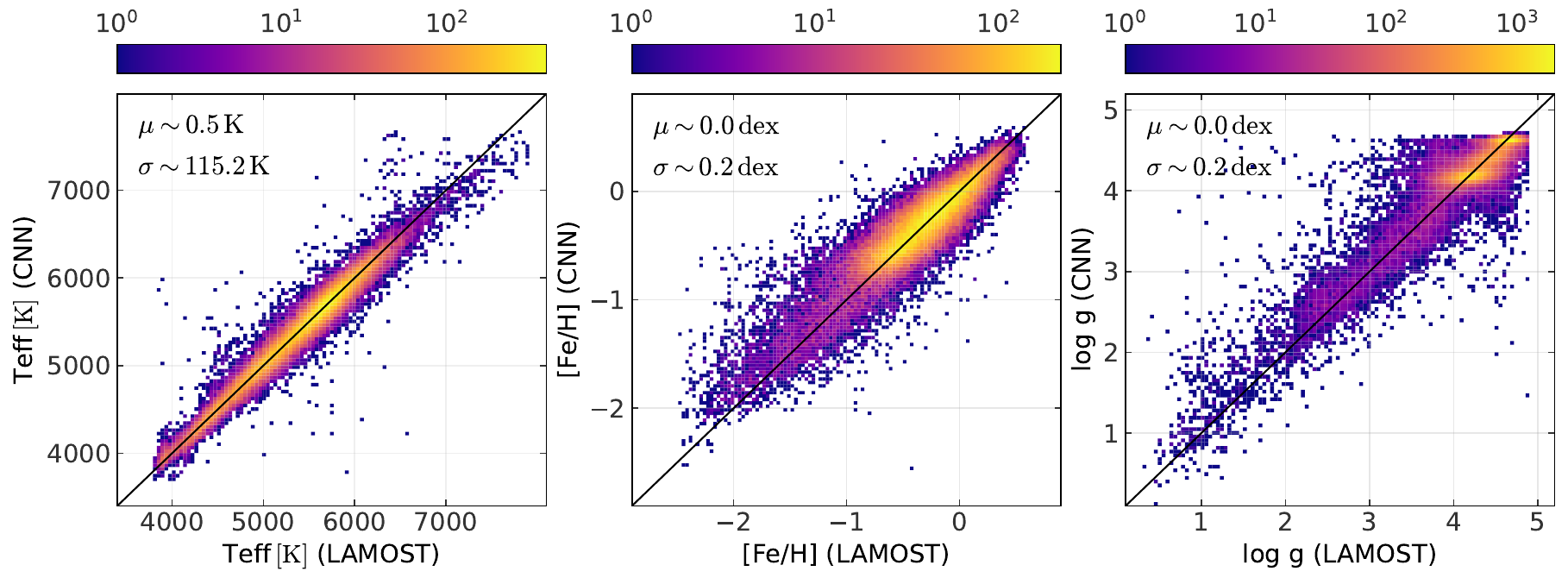}
        \caption{Comparison between the stellar parameters predicted by the CNN and the parameters given by the LAMOST catalogue. The colour scale represents the density of points.}
        \label{fig:param_val}
\end{figure*}

The horizontal axes of the three panels in Fig. \ref{fig:param_val} are the values of $T_\mathrm{eff}$, [Fe/H], and $\log\,g$ provided by LAMOST for the test sample, while the vertical axes show the corresponding values predicted by the CNN model. Overall, the stellar parameters show a good agreement. The mean values of the offsets are close to zero in all the three cases, and the $\sigma$-scatter is 115~K, 0.2~dex and 0.2~dex, in the CNN$-$LAMOST direction, respectively.

\begin{figure*}
        \centering
        \includegraphics[width=\textwidth]{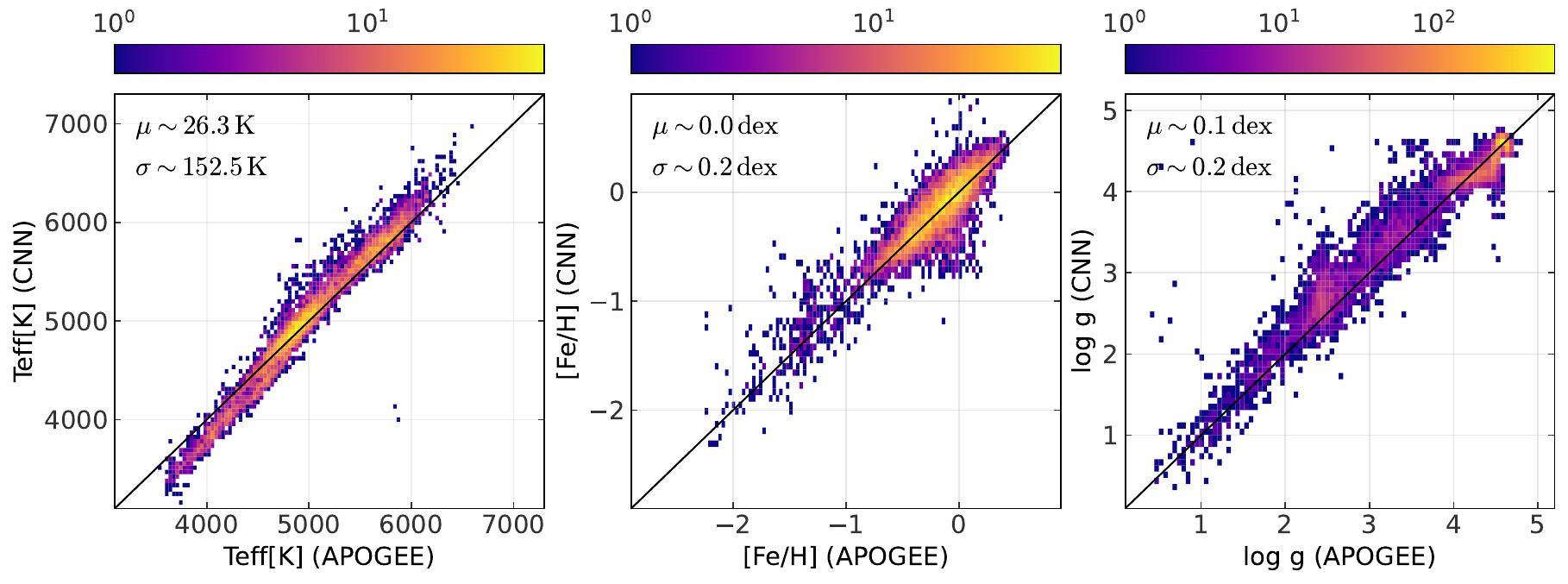}
        \caption{Comparison of the stellar parameters predicted by the CNN ans the parameters given by the APOGEE catalogue. The colour scale represents the density of points.} 
        \label{fig:val_apogee}
\end{figure*}

Additionally, in order to test the CNN with an external catalogue, we apply the model to $\sim 8\,000$ stars that are found in common between the S-PLUS iDR4 and the APOGEE DR17 catalogues. The comparison of the parameters given by the CNN and those given by APOGEE is illustrated in Fig.~\ref{fig:val_apogee}. In general, there is a good agreement between the parameters. However, for $T_\mathrm{eff} \lesssim 4\,500$~K and $T_\mathrm{eff} \gtrsim 5\,500$~K, there is a systematic offset. For [Fe/H] the agreement is good, but with higher dispersion for $\mathrm{[Fe/H] \lesssim -1}$. For $\log\,g$, the agreement is good over the whole range, although there is a slight systematic offset.
Probably, an important contribution to the observed systematic offsets is related to the systematic effects between APOGEE and LAMOST \citep[a discussion on these discrepancies found in previous data releases is given in][]{2018A&A...620A..76A}. 

\begin{table}
\centering
\caption{Stellar parameters derived with our CNN and with SPHINX}
\label{tab:stellar_parameters_splus}
\resizebox*{4.4in}{!}{%
\begin{tabular}{lccclcc}
\hline
\multicolumn{1}{c}{\textit{Gaia} DR3 ID} & \multicolumn{3}{c}{CNN (This work)}                             &  & \multicolumn{2}{c}{SPHINX}               \\ \cline{2-4} \cline{6-7} 
                                & $T_\mathrm{eff}$       & $\mathrm{[Fe/H]}$ & $\mathrm{\log\,g}$ &  & $T_\mathrm{eff}$     & $\mathrm{[Fe/H]}$ \\
                                & $\mathrm{\pm 115 (K)}$ & $\mathrm{\pm 0.2 (dex)}$   & $\mathrm{\pm 0.2 (dex)}$    &  & (K)        & (dex)     \\ \cline{1-4} \cline{6-7} 
2320869248652782208             & 6263.51                & -1.64             & 4.18               &  & 6326.93 $\pm$  18.23 & -1.92 $\pm$  0.15 \\ 
2652780576075420544             & 6351.35                & -1.68             & 4.15               &  & 6456.07 $\pm$  10.49 & -1.96 $\pm$  0.04 \\
2688149803556948992             & 5405.87                & -1.37             & 3.39               &  & 5438.38 $\pm$  23.55 & -1.61 $\pm$  0.14 \\ 
2688509099045891200             & 6289.65                & -1.42             & 4.07               &  & 6399.75 $\pm$  11.01 & -1.87 $\pm$  0.19 \\
2690227738799604224             & 5474.32                & -2.28             & 2.37               &  & 5324.79 $\pm$  23.90 & -2.34 $\pm$  0.02 \\
3504215064633606912             & 5340.02                & -1.81             & 2.45               &  & 5275.05 $\pm$  13.07 & -1.83 $\pm$  0.16 \\
3504370095773053696             & 5362.34                & -1.73             & 2.89               &  & 5334.44 $\pm$   8.11 & -1.75 $\pm$  0.23 \\
3525665333900712832             & 5745.35                & -1.71             & 3.23               &  & 5677.39 $\pm$  41.87 & -2.28 $\pm$  0.63 \\
3545705651305437568             & 5045.19                & -1.00             & 2.82               &  & 5127.88 $\pm$  28.04 & -1.28 $\pm$  0.19 \\
3546567427903076736             & 5227.47                & -1.47             & 2.72               &  & 5199.28 $\pm$  22.01 & -1.62 $\pm$  0.08 \\
3612818501033473664             & 4957.50                & -0.84             & 2.70               &  & 5039.38 $\pm$  12.45 & -1.08 $\pm$  0.19 \\
3618819154461342592             & 5992.15                & -1.12             & 4.14               &  & 6005.67 $\pm$  20.62 & -1.22 $\pm$  0.40 \\
3619582352969797888             & 4649.46                & -0.94             & 1.98               &  & 4828.08 $\pm$  74.73 & -1.16 $\pm$  0.02 \\
3620532502815040768             & 5015.09                & -0.98             & 2.56               &  & 5025.69 $\pm$  12.48 & -1.13 $\pm$  0.36 \\
3630403952568941312             & 6047.66                & -1.16             & 3.52               &  & 6044.88 $\pm$  23.88 & -1.55 $\pm$  0.59 \\ 
3630462432843699200             & 5127.26                & -1.74             & 1.48               &  & 4833.52 $\pm$  35.60 & -1.85 $\pm$  0.30 \\ 
3630970956971494912             & 5706.65                & -1.47             & 2.79               &  & 5563.23 $\pm$  50.52 & -1.72 $\pm$  0.43 \\
3631125674577631104             & 5316.31                & -1.68             & 2.58               &  & 5221.12 $\pm$  28.16 & -1.62 $\pm$  0.10 \\
3657056526743053952             & 5396.77                & -1.61             & 2.73               &  & 5318.06 $\pm$  11.42 & -1.55 $\pm$  0.12 \\ 
3696393857329932672             & 6415.20                & -0.70             & 4.16               &  & 6591.57 $\pm$  21.93 & -1.43 $\pm$  0.60 \\ 
3750233780645824768             & 6241.03                & -1.78             & 4.14               &  & 6283.97 $\pm$   1.23 & -1.87 $\pm$  0.21 \\
3752558526183725440             & 5480.49                & -0.89             & 3.63               &  & 5576.75 $\pm$  60.16 & -1.11 $\pm$  0.13 \\
3753526229559762176             & 5663.39                & -1.78             & 3.34               &  & 5665.44 $\pm$  28.71 & -1.92 $\pm$  0.10 \\ 
3772348601612372992             & 5610.63                & -1.25             & 2.69               &  & 5483.33 $\pm$  94.22 & -1.41 $\pm$  0.33 \\
3793377208170393984             & 5774.15                & -1.21             & 3.22               &  & 5985.35 $\pm$  41.78 & -0.91 $\pm$  0.33 \\ 
3891412241883772928             & 5163.16                & -1.57             & 2.08               &  & 5031.82 $\pm$ 132.40 & -1.85 $\pm$  0.11 \\ 
4227521019370523648             & 5146.82                & -0.85             & 2.77               &  & 5160.48 $\pm$  41.67 & -1.22 $\pm$  0.23 \\
4229238598267973376             & 4524.20                & -0.92             & 1.72               &  & 4621.65 $\pm$  26.24 & -1.12 $\pm$  0.17 \\
4235951322918940928             & 4110.82                & -0.35             & 1.29               &  & 4324.94              & -0.91             \\
4236552309105405824             & 5067.28                & -1.05             & 2.88               &  & 5149.85 $\pm$  18.75 & -1.06 $\pm$  0.15 \\
4242563614044957696             & 5736.21                & -0.23             & 4.38               &  & 5669.36 $\pm$  52.96 & -0.34 $\pm$  0.19 \\
4681680795508995840             & 6163.68                & -1.41             & 4.18               &  & 6181.12 $\pm$  11.74 & -1.84 $\pm$  0.31 \\
4787052870314603520             & 5492.13                & -1.45             & 3.43               &  & 5469.04 $\pm$  38.04 & -1.65 $\pm$  0.15 \\
4788482544668341632             & 5931.65                & -1.32             & 3.68               &  & 5978.14 $\pm$  59.86 & -1.92 $\pm$  0.58 \\
4788911315546547328             & 5803.39                & -1.56             & 3.04               &  & 5727.32 $\pm$  37.99 & -1.72 $\pm$  0.62 \\
4790354875531489024             & 6034.51                & -2.20             & 4.09               &  & 6043.32 $\pm$   3.44 & -2.19             \\
4856695936817617280             & 5150.69                & -0.48             & 3.17               &  & 5179.08 $\pm$   6.40 & -0.42 $\pm$  0.09 \\
4882090566610735360             & 5547.61                & -1.64             & 3.09               &  & 5464.47 $\pm$  36.85 & -1.88 $\pm$  0.22 \\
5048231102894377344             & 5397.33                & -1.92             & 2.52               &  & 5306.82 $\pm$  18.34 & -1.80 $\pm$  0.23 \\
5401875170994688896             & 5923.29                & -0.78             & 3.58               &  & 5834.46 $\pm$ 177.77 & -0.28 $\pm$  0.29 \\
5441944638890204160             & 5092.00                & -0.17             & 4.64               &  & 5189.06 $\pm$  21.25 & -0.15 $\pm$  0.08 \\
5463127795547217024             & 5193.75                & -0.66             & 4.57               &  & 5253.21 $\pm$   2.11 & -1.15 $\pm$  0.29 \\
5469865053045546112             & 5364.07                & -1.22             & 3.12               &  & 5371.00 $\pm$  23.89 & -1.36 $\pm$  0.09 \\
5473218563510860032             & 3366.22                &  0.79             & 4.49               &  & 3943.43              & -2.34             \\
5475847186574041984             & 4098.79                & -0.28             & 4.57               &  & 4170.70 $\pm$  17.39 & -1.63             \\
5674511482965847168             & 3733.44                & -0.65             & 4.60               &  & 3991.22              & -2.37             \\
6270952693712299008             & 5508.00                & -1.18             & 3.40               &  & 5525.74 $\pm$   6.50 & -1.27 $\pm$  0.15 \\
6478130959611566720             & 6048.57                & -1.62             & 4.04               &  & 6178.14 $\pm$  65.52 & -2.06 $\pm$  0.23 \\ 
6478219813895286912             & 5538.65                & -2.00             & 2.76               &  & 5320.81 $\pm$  46.98 & -2.35 $\pm$  0.15 \\
6503261122497594624             & 5060.01                & -1.54             & 1.96               &  & 4928.41 $\pm$   6.11 & -1.49 $\pm$  0.15 \\
6575967187877774720             & 4154.52                & -0.84             & 1.46               &  & 4476.64 $\pm$  76.90 & -1.26             \\ 
6580745253096436224             & 5146.67                & -1.90             & 2.11               &  & 4993.92 $\pm$  17.63 & -2.05 $\pm$  0.16 \\
6617426988221664640             & 5066.87                & -1.64             & 2.40               &  & 4992.29 $\pm$  11.21 & -1.70 $\pm$  0.11 \\
6619340275892884224             & 4989.84                & -1.20             & 2.11               &  & 4979.87 $\pm$  67.02 & -1.27 $\pm$  0.29 \\
6625197335678814208             & 5597.84                & -1.79             & 2.74               &  & 5457.47 $\pm$  35.24 & -2.09 $\pm$  0.09 \\
6775627229489106688             & 6238.07                & -1.05             & 4.07               &  & 6374.05 $\pm$  15.49 & -1.63 $\pm$  0.33 \\
6776360157068155520             & 4724.03                & -1.01             & 1.98               &  & 4790.26 $\pm$  24.17 & -1.12 $\pm$  0.05 \\
6782909879114518016             & 5184.62                & -1.35             & 2.67               &  & 5144.72 $\pm$  25.51 & -1.52 $\pm$  0.08 \\
6785190193214463872             & 6010.64                & -1.20             & 3.87               &  & 6061.83 $\pm$  43.96 & -1.74 $\pm$  0.46 \\
6785833613675654912             & 5592.85                & -1.85             & 3.27               &  & 5555.44 $\pm$  24.52 & -1.98 $\pm$  0.26 \\
6790317563828425728             & 4870.09                & -0.84             & 2.47               &  & 4955.32 $\pm$  33.30 & -1.01 $\pm$  0.15 \\
6790795542148925952             & 5040.45                & -0.98             & 2.85               &  & 5129.11 $\pm$  16.11 & -1.07 $\pm$  0.07 \\
6812361977394894464             & 5748.10                & -1.86             & 3.21               &  & 5646.10 $\pm$  34.74 & -2.14 $\pm$  0.18 \\
6917206978109246080             & 5792.61                & -0.05             & 4.23               &  & 5858.17 $\pm$  11.82 & -0.18 $\pm$  0.13 \\ \hline
\end{tabular}%
}
\end{table}

\subsection{Comparison with SPHINX}

Here we compare the results of the CNN to the results obtained with the SPHINX artificial neural network \citep{2019A&A...622A.182W}. The current model of SPHINX \citep{2021ApJ...912..147W} has been trained using colours and magnitudes from the 12 filter photometry of S-PLUS. In particular, the model applied here is based on the S-PLUS iDR4. The training labels were taken from a combination of LAMOST DR7 and the SEGUE DR9 \citep[Sloan Extension for Galactic Understanding and Exploration,][]{2009AJ....137.4377Y}\footnote{\url{https://www.sdss4.org/surveys/segue/}}, totalling $\sim7\,000$ stars. The network was trained for a range of $4\,000\leq T_\mathrm{eff}\leq 6\,750$~K and $-2.5 \leq\mathrm{[Fe/H]}/\mathrm{dex} \leq 0.0$, and it only provides estimates for $T_\mathrm{eff}$ and [Fe/H]. 

\begin{figure}
    \centering
    \includegraphics[width=\columnwidth]{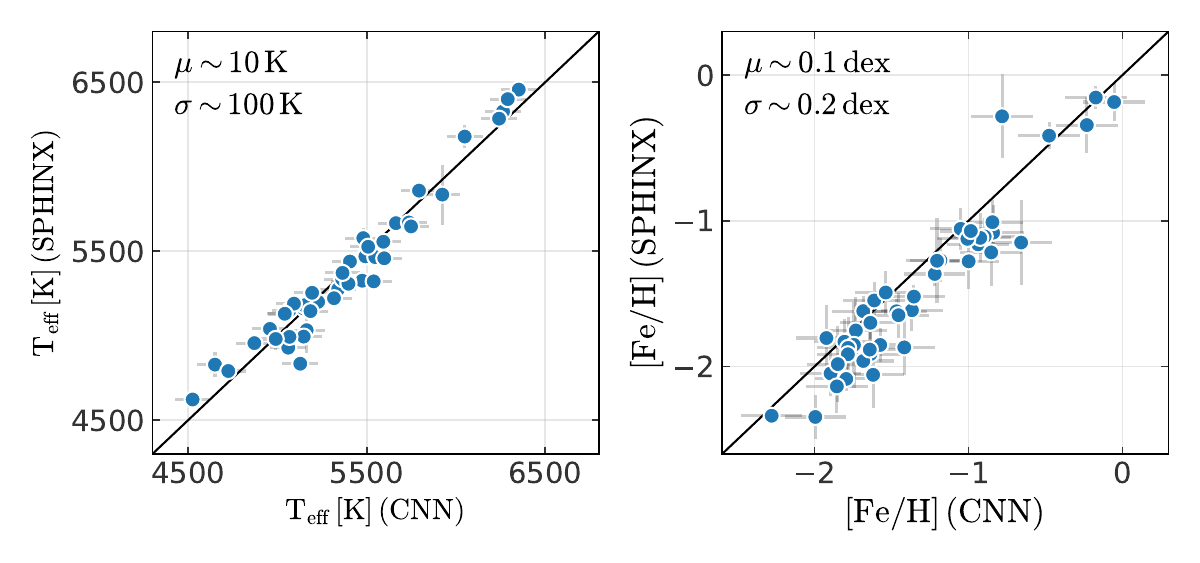}
    \caption{Comparison between the effective temperature (left panel) and the metallicity (right panel) using the CNN and SPHINX.}
    \label{fig:cnn_sphinx}
\end{figure}

The application of the current SPHINX model to our HiVel sample produces the results reported in Table \ref{tab:stellar_parameters_splus}. The comparison with the CNN-estimated values is shown in Fig.~\ref{fig:cnn_sphinx}, for sources with metallicity uncertainties $<0.3$~dex and with uncertainty information in SPHINX. In the case of the effective temperature (left panel), we see that there is a slight positive offset of 10~K, with a dispersion of 100~K. In the case of the [Fe/H] (right panel), we clearly see a systematic offset of 0.1~dex, and a dispersion of 0.2~dex in the CNN$-$SPHINX direction. 

\subsection{The Hertzsprung–Russell diagram}

\begin{figure}
        \centering
        \includegraphics[width=\columnwidth]{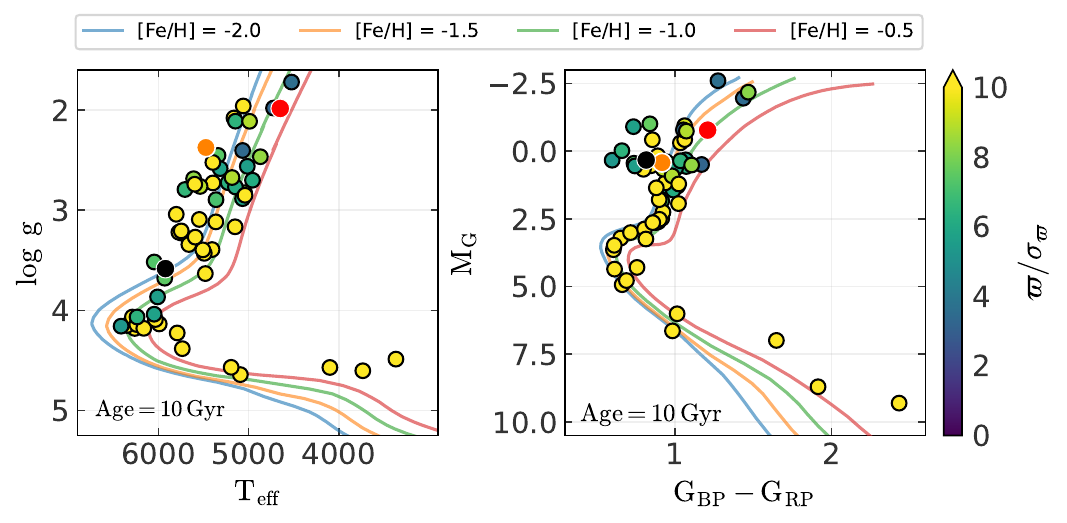}
        \caption{Location of the high velocity stars in the Kiel diagram (left), using the CNN values, and in the colour-magnitude diagram (right), using the \textit{Gaia} magnitudes. The colour lines are isochrones computed for different metallicities (see text for details). The only unbound candidate star (\textit{Gaia} DR3 2690227738799604224) is shown in orange. The two stars with a probable origin in the GC, identified through kinematic analysis, are denoted by a black dot (\textit{Gaia} DR3 5401875170994688896) and a red dot (\textit{Gaia} DR3 3619582352969797888). The colour code is the same as in Fig.~\ref{fig:scape_velocity_curve}.}
        \label{fig:HR_diagram}
\end{figure}

In Fig.~\ref{fig:HR_diagram}, we plot the Kiel diagram (left panel) using the CNN-estimated stellar parameters reported in Table~\ref{tab:stellar_parameters_splus}, and the CMD diagram (right panel) using \textit{Gaia} photometry. These diagrams aim to identify the evolutionary stage of the HiVel stars. The absolute magnitude in the CMD is calculated using the G-band apparent magnitudes, corrected by extinction using the SFD2D dust map \citep{Schlafly2011}, through the \texttt{dustmaps} package \citep{Green2018} and the distances estimated by \citet{2021yCat.1352....0B}. The curves in blue, orange, green and red are the isochrones computed for metallicities $-2.0$, $-1.5$, $-1.0$ and $-0.5$~dex, respectively. The data for the isochrones have been generated using \texttt{PARSEC} \citep[PAdova and TRieste Stellar Evolution Code,][]{Bressan2012}\footnote{\url{http://stev.oapd.inaf.it/cmd/}}. 

In both diagrams, most of the stars in the HiVel sample are in the red giant branch, some stars are in the turn-off region, and a few are in the low-mass main sequence. The observed effective temperature range in the sample, from $\sim3\,500$~K to $\sim6\,500$~K, and their location in the H-R diagram, disfavour the presence of runaway or white dwarf hyper-runaway stars (stars with effective temperatures higher than $\sim 7\,000$ K) in the sample. 

Additionally, considering the stellar masses provided by the Gaia EDR3 StarHorse catalog, based in photometric information \citep{2022A&A...658A..91A}, we calculate the corresponding stellar radii using: (i) the values of $\log\,g$ from the StarHorse catalogue itself (SH radius), and (ii) the values of $\log\,g$ predicted by our CNN model (CNN radius). In both calculations, the masses are taken from the StarHorse catalog. The resulting radii are reported in Table \ref{tab:mass_logg_rad}, and most of them are consistent with their position in the Kiel diagram.

The fact that some stars are inconsistent with the corresponding isochrones shown in Fig. \ref{fig:HR_diagram} may be related to the large dispersion in the inferred $\log g$ values.

\begin{table}
\centering
\caption{Stellar radii of the high velocity stars.}
\label{tab:mass_logg_rad}
\resizebox*{4.2in}{!}{%
\begin{tabular}{lccccc}
\hline
\multicolumn{1}{c}{\textit{Gaia} DR3 ID} & Mass (SH)                 & Log g (SH)                & Radius (SH)$^\mathrm{a}$            & Log g (CNN)              & Radius (CNN)$^\mathrm{b}$           \\
                                & $\mathrm{(M_{\odot})}$    & $\mathrm{(dex)}$          & $\mathrm{(R_{\odot})}$ & $\mathrm{\pm 0.2 (dex)}$ & $\mathrm{(R_{\odot})}$ \\ \hline
2320869248652782208   &  $0.86_{-0.06}^{+0.04}$  &  $3.90_{-0.06}^{+0.01}$  &  $1.73_{-0.04}^{+0.07}$ &  4.18  &  $1.25_{-0.25}^{+0.28}$ \\
2652780576075420544   &  $0.94_{-0.04}^{+0.17}$  &  $4.11_{-0.02}^{+0.10}$  &  $1.41_{-0.06}^{+0.04}$ &  4.15  &  $1.41_{-0.25}^{+0.27}$ \\
2688149803556948992*  &  $0.90_{-0.00}^{+0.00}$  &  $3.35_{-0.16}^{+0.00}$  &  $3.39_{-0.08}^{+0.60}$ &  3.39  &  $3.23_{-0.66}^{+0.85}$ \\
2688509099045891200   &  $0.85_{-0.02}^{+0.20}$  &  $4.01_{-0.03}^{+0.13}$  &  $1.53_{-0.11}^{+0.06}$ &  4.07  &  $1.48_{-0.27}^{+0.32}$ \\
2690227738799604224   &  $0.80_{-0.02}^{+0.06}$  &  $2.45_{-0.04}^{+0.22}$  &  $8.75_{-1.83}^{+0.38}$ &  2.37  &  $9.84_{-1.86}^{+2.25}$ \\
3504215064633606912   &  $0.82_{-0.04}^{+0.02}$  &  $2.38_{-0.09}^{+0.04}$  &  $9.74_{-0.35}^{+0.86}$ &  2.45  &  $8.94_{-1.70}^{+2.28}$ \\
3504370095773053696   &  $0.90_{-0.06}^{+0.01}$  &  $3.00_{-0.15}^{+0.19}$  &  $4.97_{-0.90}^{+0.86}$ &  2.89  &  $5.63_{-1.02}^{+1.36}$ \\
3525665333900712832*  &  $0.73_{-0.00}^{+0.00}$  &  $1.91_{-0.00}^{+0.26}$  &  $15.6_{-4.25}^{+0.02}$ &  3.23  &  $3.52_{-0.76}^{+0.77}$ \\
3545705651305437568   &  $0.85_{-0.04}^{+0.04}$  &  $2.68_{-0.25}^{+0.14}$  &  $7.12_{-1.05}^{+2.19}$ &  2.82  &  $5.98_{-1.15}^{+1.47}$ \\
3546567427903076736   &  $0.82_{-0.04}^{+0.04}$  &  $2.44_{-0.03}^{+0.24}$  &  $8.92_{-1.82}^{+0.34}$ &  2.72  &  $6.57_{-1.19}^{+1.59}$ \\
3612818501033473664   &  $0.91_{-0.05}^{+0.00}$  &  $2.44_{-0.14}^{+0.16}$  &  $9.48_{-1.44}^{+1.48}$ &  2.70  &  $7.07_{-1.51}^{+1.71}$ \\
3618819154461342592*  &  $0.90_{-0.02}^{+0.01}$  &  $3.78_{-0.05}^{+0.03}$  &  $2.02_{-0.05}^{+0.11}$ &  4.14  &  $1.37_{-0.28}^{+0.31}$ \\
3619582352969797888   &  $0.86_{-0.02}^{+0.05}$  &  $1.59_{-0.12}^{+0.34}$  &  $24.8_{-6.59}^{+3.40}$ &  1.98  &  $16.0_{-3.10}^{+3.83}$ \\
3620532502815040768   &  $0.81_{-0.01}^{+0.07}$  &  $2.34_{-0.28}^{+0.07}$  &  $10.5_{-0.87}^{+3.18}$ &  2.56  &  $8.03_{-1.57}^{+1.89}$ \\
3630403952568941312   &  $0.73_{-0.00}^{+0.02}$  &  $2.58_{-0.10}^{+0.06}$  &  $7.37_{-0.50}^{+0.70}$ &  3.52  &  $2.53_{-0.55}^{+0.62}$ \\
3630462432843699200   &  $0.79_{-0.01}^{+0.01}$  &  $1.75_{-0.09}^{+0.03}$  &  $19.8_{-0.81}^{+2.07}$ &  1.48  &  $27.4_{-5.53}^{+7.03}$ \\
3630970956971494912*  &  $0.74_{-0.00}^{+0.00}$  &  $2.49_{-0.00}^{+0.00}$  &  $8.14_{-0.02}^{+0.00}$ &  2.79  &  $5.92_{-1.31}^{+1.35}$ \\
3631125674577631104   &  $0.80_{-0.04}^{+0.05}$  &  $2.47_{-0.03}^{+0.46}$  &  $8.53_{-3.00}^{+0.28}$ &  2.58  &  $7.67_{-1.43}^{+1.68}$ \\
3657056526743053952   &  $0.81_{-0.01}^{+0.06}$  &  $3.17_{-0.06}^{+0.08}$  &  $3.94_{-0.29}^{+0.21}$ &  2.73  &  $6.65_{-1.22}^{+1.51}$ \\
3696393857329932672*  &  $0.79_{-0.06}^{+0.00}$  &  $2.75_{-0.24}^{+0.02}$  &  $6.20_{-0.15}^{+1.87}$ &  4.16  &  $1.22_{-0.24}^{+0.31}$ \\
3750233780645824768   &  $0.83_{-0.00}^{+0.07}$  &  $4.35_{-0.00}^{+0.04}$  &  $1.00_{-0.01}^{+0.01}$ &  4.14  &  $1.32_{-0.26}^{+0.30}$ \\
3752558526183725440   &  $1.15_{-0.31}^{+0.13}$  &  $3.54_{-0.17}^{+0.05}$  &  $3.05_{-0.24}^{+0.29}$ &  3.63  &  $2.62_{-0.46}^{+0.55}$ \\
3753526229559762176   &  $0.88_{-0.02}^{+0.02}$  &  $3.55_{-0.04}^{+0.00}$  &  $2.62_{-0.03}^{+0.07}$ &  3.34  &  $3.39_{-0.68}^{+0.86}$ \\
3772348601612372992   &  $0.81_{-0.05}^{+0.02}$  &  $2.44_{-0.03}^{+0.02}$  &  $8.98_{-0.20}^{+0.21}$ &  2.69  &  $6.81_{-1.20}^{+1.80}$ \\
3793377208170393984   &  $0.81_{-0.07}^{+0.03}$  &  $2.47_{-0.03}^{+0.17}$  &  $8.47_{-1.07}^{+0.36}$ &  3.22  &  $3.67_{-0.66}^{+0.76}$ \\
3891412241883772928   &  $0.79_{-0.05}^{+0.06}$  &  $2.00_{-0.06}^{+0.08}$  &  $14.7_{-0.90}^{+0.73}$ &  2.08  &  $13.6_{-2.46}^{+2.93}$ \\
4227521019370523648   &  $0.83_{-0.04}^{+0.06}$  &  $2.77_{-0.29}^{+0.23}$  &  $6.31_{-1.30}^{+2.26}$ &  2.77  &  $6.33_{-1.23}^{+1.50}$ \\
4229238598267973376   &  $0.78_{-0.05}^{+0.05}$  &  $1.29_{-0.19}^{+0.03}$  &  $34.1_{-1.52}^{+6.87}$ &  1.72  &  $20.4_{-3.72}^{+4.98}$ \\
4235951322918940928   &  $0.88_{-0.15}^{+0.14}$  &  $1.21_{-0.20}^{+0.12}$  &  $39.0_{-4.13}^{+7.12}$ &  1.29  &  $35.3_{-5.76}^{+7.26}$ \\
4236552309105405824   &  $0.81_{-0.02}^{+0.02}$  &  $2.43_{-0.03}^{+0.24}$  &  $9.03_{-1.85}^{+0.26}$ &  2.88  &  $5.45_{-1.04}^{+1.31}$ \\
4242563614044957696**   &  $1.05_{-0.16}^{+0.01}$  &  $3.82_{-0.03}^{+0.00}$  &  $2.09_{-0.14}^{+0.01}$ &  4.38  &  $1.07_{-0.20}^{+0.24}$ \\
4681680795508995840   &  $0.76_{-0.05}^{+0.04}$  &  $4.50_{-0.03}^{+0.03}$  &  $0.81_{-0.02}^{+0.02}$ &  4.18  &  $1.19_{-0.21}^{+0.24}$ \\
4787052870314603520   &  $0.81_{-0.03}^{+0.05}$  &  $3.41_{-0.08}^{+0.03}$  &  $3.00_{-0.10}^{+0.20}$ &  3.43  &  $2.92_{-0.54}^{+0.66}$ \\
4788482544668341632   &  $0.79_{-0.06}^{+0.01}$  &  $2.51_{-0.05}^{+0.06}$  &  $8.04_{-0.41}^{+0.38}$ &  3.68  &  $2.13_{-0.43}^{+0.48}$ \\
4788911315546547328*  &  $0.84_{-0.03}^{+0.01}$  &  $2.73_{-0.01}^{+0.00}$  &  $6.51_{-0.08}^{+0.05}$ &  3.04  &  $4.61_{-0.91}^{+1.08}$ \\
4790354875531489024   &  $0.85_{-0.08}^{+0.08}$  &  $4.49_{-0.04}^{+0.03}$  &  $0.86_{-0.03}^{+0.03}$ &  4.09  &  $1.38_{-0.25}^{+0.29}$ \\
4856695936817617280*  &  $0.94_{-0.00}^{+0.02}$  &  $3.00_{-0.01}^{+0.03}$  &  $5.06_{-0.13}^{+0.03}$ &  3.17  &  $4.31_{-0.93}^{+1.03}$ \\
4882090566610735360   &  $0.81_{-0.03}^{+0.05}$  &  $3.07_{-0.06}^{+0.01}$  &  $4.46_{-0.09}^{+0.18}$ &  3.09  &  $4.34_{-0.82}^{+1.03}$ \\
5048231102894377344   &  $0.82_{-0.04}^{+0.02}$  &  $2.86_{-0.06}^{+0.03}$  &  $5.58_{-0.14}^{+0.28}$ &  2.52  &  $8.27_{-1.57}^{+2.09}$ \\
5401875170994688896   &  $0.82_{-0.03}^{+0.02}$  &  $2.43_{-0.01}^{+0.05}$  &  $9.02_{-0.32}^{+0.14}$ &  3.58  &  $2.44_{-0.46}^{+0.67}$ \\
5441944638890204160   &  $0.74_{-0.04}^{+0.04}$  &  $4.62_{-0.02}^{+0.02}$  &  $0.70_{-0.02}^{+0.01}$ &  4.64  &  $0.69_{-0.14}^{+0.16}$ \\
5463127795547217024*  &  $0.62_{-0.01}^{+0.01}$  &  $4.75_{-0.01}^{+0.01}$  &  $0.55_{-0.00}^{+0.00}$ &  4.57  &  $0.70_{-0.16}^{+0.17}$ \\
5469865053045546112   &  $0.86_{-0.00}^{+0.23}$  &  $3.33_{-0.04}^{+0.08}$  &  $3.42_{-0.16}^{+0.18}$ &  3.12  &  $4.51_{-0.72}^{+0.88}$ \\
5473218563510860032*  &  $0.51_{-0.00}^{+0.01}$  &  $4.74_{-0.00}^{+0.00}$  &  $0.50_{-0.00}^{+0.00}$ &  4.49  &  $0.69_{-0.14}^{+0.16}$ \\
5475847186574041984*  &  $0.78_{-0.00}^{+0.01}$  &  $4.51_{-0.00}^{+0.04}$  &  $0.81_{-0.03}^{+0.00}$ &  4.57  &  $0.78_{-0.16}^{+0.19}$ \\
5674511482965847168*  &  $0.49_{-0.00}^{+0.00}$  &  $4.78_{-0.00}^{+0.00}$  &  $0.47_{-0.00}^{+0.00}$ &  4.60  &  $0.59_{-0.13}^{+0.14}$ \\
6270952693712299008   &  $0.84_{-0.06}^{+0.11}$  &  $2.90_{-0.17}^{+0.11}$  &  $5.51_{-0.56}^{+0.88}$ &  3.40  &  $3.12_{-0.56}^{+0.67}$ \\
6478130959611566720   &  $0.77_{-0.04}^{+0.16}$  &  $2.43_{-0.01}^{+0.01}$  &  $8.93_{-0.21}^{+0.71}$ &  4.04  &  $1.43_{-0.24}^{+0.29}$ \\
6478219813895286912   &  $0.80_{-0.02}^{+0.07}$  &  $2.46_{-0.04}^{+0.12}$  &  $8.69_{-0.76}^{+0.33}$ &  2.76  &  $6.32_{-1.19}^{+1.52}$ \\
6503261122497594624   &  $0.79_{-0.01}^{+0.06}$  &  $2.02_{-0.10}^{+0.01}$  &  $14.9_{-0.43}^{+1.21}$ &  1.96  &  $16.0_{-2.58}^{+3.68}$ \\
6575967187877774720   &  $0.79_{-0.02}^{+0.07}$  &  $1.28_{-0.04}^{+0.05}$  &  $34.2_{-1.16}^{+1.26}$ &  1.46  &  $28.0_{-4.80}^{+6.54}$ \\
6580745253096436224   &  $0.85_{-0.04}^{+0.02}$  &  $2.40_{-0.03}^{+0.02}$  &  $9.60_{-0.16}^{+0.24}$ &  2.11  &  $13.5_{-2.72}^{+3.51}$ \\
6617426988221664640   &  $0.84_{-0.06}^{+0.01}$  &  $2.00_{-0.15}^{+0.03}$  &  $15.1_{-0.48}^{+2.61}$ &  2.40  &  $9.55_{-1.84}^{+2.25}$ \\
6619340275892884224   &  $0.80_{-0.02}^{+0.03}$  &  $1.82_{-0.19}^{+0.06}$  &  $18.7_{-1.32}^{+4.13}$ &  2.11  &  $13.3_{-2.47}^{+3.02}$ \\
6625197335678814208   &  $0.82_{-0.04}^{+0.02}$  &  $2.41_{-0.03}^{+0.02}$  &  $9.35_{-0.17}^{+0.20}$ &  2.74  &  $6.46_{-1.30}^{+1.46}$ \\
6775627229489106688   &  $0.79_{-0.06}^{+0.00}$  &  $2.57_{-0.06}^{+0.06}$  &  $7.59_{-0.35}^{+0.41}$ &  4.07  &  $1.36_{-0.25}^{+0.33}$ \\
6776360157068155520   &  $0.88_{-0.08}^{+0.12}$  &  $2.22_{-0.41}^{+0.13}$  &  $12.7_{-1.75}^{+7.13}$ &  1.98  &  $16.2_{-2.70}^{+2.97}$ \\
6782909879114518016   &  $0.82_{-0.03}^{+0.08}$  &  $2.67_{-0.24}^{+0.12}$  &  $7.17_{-0.87}^{+1.91}$ &  2.67  &  $7.07_{-1.26}^{+1.63}$ \\
6785190193214463872   &  $0.73_{-0.00}^{+0.10}$  &  $2.62_{-0.12}^{+0.06}$  &  $7.19_{-0.42}^{+0.72}$ &  3.87  &  $1.72_{-0.32}^{+0.37}$ \\
6785833613675654912   &  $0.86_{-0.07}^{+0.04}$  &  $3.40_{-0.06}^{+0.03}$  &  $3.05_{-0.10}^{+0.14}$ &  3.27  &  $3.57_{-0.63}^{+0.82}$ \\
6790317563828425728   &  $0.86_{-0.08}^{+0.04}$  &  $2.41_{-0.03}^{+0.01}$  &  $9.56_{-0.26}^{+0.20}$ &  2.47  &  $9.00_{-1.54}^{+2.17}$ \\
6790795542148925952   &  $0.91_{-0.03}^{+0.03}$  &  $2.71_{-0.04}^{+0.09}$  &  $6.97_{-0.58}^{+0.26}$ &  2.85  &  $6.03_{-1.24}^{+1.29}$ \\
6812361977394894464   &  $0.78_{-0.04}^{+0.11}$  &  $2.20_{-0.08}^{+0.07}$  &  $11.9_{-0.65}^{+0.82}$ &  3.21  &  $3.74_{-0.68}^{+0.87}$ \\
6917206978109246080   &  $0.93_{-0.01}^{+0.03}$  &  $4.30_{-0.03}^{+0.04}$  &  $1.14_{-0.04}^{+0.03}$ &  4.23  &  $1.26_{-0.24}^{+0.28}$ \\ \hline
\multicolumn{6}{l}{$^\mathrm{a}$ Calculated using $\log g$ from the \textit{Gaia} EDR3 StarHorse catalog} \\
\multicolumn{6}{l}{$^\mathrm{b}$ Calculated using $\log g$ determined by the CNN} \\
\multicolumn{6}{l}{* Sources with flag activated $\mathrm{(sh\_outflag \neq 0)}$ in StarHorse} \\
\multicolumn{6}{l}{** Source classified as variable star in \textit{Gaia} DR3} 
\end{tabular}%
}
\end{table}

\subsection{Chemical constraints}\label{sect:chemical}

Accurate information on the chemical composition of the stars' atmospheres is essential to constraint the birthplaces of high velocity stars through \textit{chemical tagging}. In the absence of such information for the HiVel sample in this study, restrictions on their origin will be based on their [Fe/H] and [Mg/Fe] ratio, derived from the photometric S-PLUS data, and on complementary chemical information found in spectroscopic surveys (see Table~\ref{tab:stellar_abundances}).

Based on the CNN metallicity, we have identified 11 stars with $\mathrm{[Fe/H]} \geq -0.8$ that could be associated with either the thick or the thin disc populations. Given that [Mg/Fe] information allows these populations to be disentangled \citep{1998A&A...338..161F, 2017AJ....154...94M}, and that it is one of the most accurate elements measured by APOGEE \citep{2018AJ....156..126J}, we trained an Artificial Neural Network (ANN) with data common to APOGEE DR17 and SPLUS iDR4 ($\sim 8\,000$ stars) to get $T_\mathrm{eff}$, [Fe/H], and [Mg/Fe] for stars with $\mathrm{[Fe/H] > -1}$. We consider this interval because it includes more data in common between the catalogues and shows a lower metallicity dispersion. On the other hand, we used an ANN in this case because it was more efficient for this sample, but since there are several hyperparameters involved in the training process, further analysis is needed.

The one-to-one comparison between the predicted values using the ANN and those given by APOGEE, for the test sample ($20\,\%$ of the data), is shown in the left panel of Fig.~\ref{fig:mg_distribution}. We verify that the mean of the differences is $\sim 0$ dex, and the scatter is $\sim 0.07$ dex, showing a good agreement in general. Nevertheless, there are systematic offsets for values [Mg/Fe] either lower or higher than $\sim 0.2$~dex. The middle panel in Fig.~\ref{fig:mg_distribution} shows the density plot of the [Mg/Fe] vs. {Fe/H] distribution predicted by the ANN for the test sample. The right panel shows density plot of the [Mg/Fe] vs. [Fe/H] distribution predicted by the ANN for the APOGEE data. The dashed lines in these panels indicate the approximate boundary between the thin disc (lower region) and the thick disc (upper region). The right panel also shows the distribution of the HiVel stars with [Fe/H] > -1, as predicted by the ANN. Most of the HiVel stars are located in the upper-left part of the figure, along to the not well defined border between the thick disc and the halo. The red and black circles represent the stars \textit{Gaia} DR3 5401875170994688896 and \textit{Gaia} DR3 3619582352969797888, respectively, that might have a chance of being originated in the centre of the Galaxy. The red circle is in the transition region between the thick disc and the halo, so the chance of having an origin in the GC is very low. The black circle is located in the thin disc region, which also overlaps with the expected region for the bulge stars \citep{2018ARA&A..56..223B, 2021A&A...656A.156Q, 2023A&A...671A..94N}, favouring its origin in the central part of the MW. 

On the other hand, according to their position in the [Mg/Fe] vs. [Fe/H] diagram, four stars are located between the thin disc and the halo. Two stars are in the highest density region, with a chance of being thin disc stars. This, in addition to the kinematic analysis, would imply in low mass disc stars with halo kinematics, which is something unexpected. Analysing the stellar properties, we see that in the H-R diagram these stars are located in the low mass main-sequence, with a G apparent magnitude around 15 and low signal to noise (SNR < 5). Therefore, it is possible that these stars have a spurious radial velocity, and consequently a fake halo star kinematic behaviour. However, a spectroscopic study would be required to confirm the radial velocities given by \textit{Gaia}.

\begin{figure}
        \centering
        \includegraphics[width=\columnwidth]{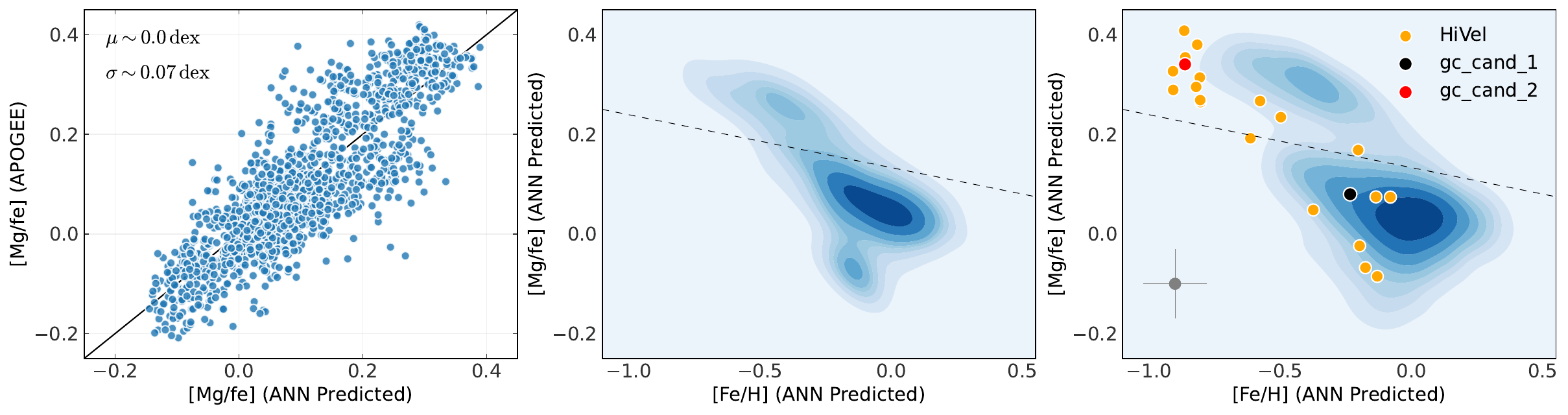}
        \caption{The left panel shows the comparison between the [Mg/Fe] ratio predicted by the ANN and the ratio given by APOGEE, for the test sample. The middle panel shows the density plot of the [Mg/Fe] vs. [Fe/H] distribution of the test sample, predicted by ANN. The right panel shows the [Mg/Fe] vs. [Fe/H] distribution, predicted by the ANN, for the APOGEE sample (density plot) and for the HiVel sample (dots). Stars with probable origin in the GC are denoted by a black dot (\textit{Gaia} DR3 5401875170994688896) and a red dot (\textit{Gaia} DR3 3619582352969797888). The dashed lines indicate the approximate boundary between the thin disc (lower region) and the thick disc (upper region).}    
        \label{fig:mg_distribution}
\end{figure}

The remaining stars in our sample, with $\mathrm{[Fe/H]} < -1$, can be classified as metal-poor stars, which is a characteristic expected of halo stars. Knowledge of $\alpha$-element abundances is essential to determine whether halo stars in the $-1.8 < \mathrm{[Fe/H]} < -0.8$~dex metallicity range have an extra-galactic origin due to the merger of extra-galactic stellar systems with the MW (so called accreted stars, characterised by a low $\alpha$-elements abundance), or if they formed locally within the MW (so called in situ stars, characterised by a high $\alpha$-elements abundance). The limit between accreted and in situ stars is approximately given by $\mathrm{[\alpha/Fe]} = -0.15\,\mathrm{[Fe/H]} + 0.07$ \citep{Nissen2010}. Using this limit together with the $\alpha$-element values reported in Table~\ref{tab:stellar_abundances}, most of the HiVel stars fall in the region of accreted stars. Note, however, that this limit may be slightly off due to systematic differences between the spectra used by \citet{Nissen2010} and those from other spectroscopic surveys.

It is important to note that the analysis presented in this section is based on results obtained using machine learning algorithms and medium/low resolution spectroscopic surveys. Therefore, to confirm the results, it is mandatory to perform a chemical abundance analysis using high resolution spectra.

\begin{table}
\centering
\caption{Metallicity, $\alpha$-elements and [Mg/Fe] abundance from spectroscopic surveys.}
\label{tab:stellar_abundances}
\begin{tabular}{lcccl}
\hline
\multicolumn{1}{c}{\textit{Gaia} DR3 ID} &  [Fe/H]   & [$\alpha$/Fe]    &       [Mg/Fe]  & \multicolumn{1}{c}{Survey}      \\
\hline        
3630403952568941312  & -1.21  $\pm$  0.21  &  0.26  $\pm$   0.07  &   0.14 $\pm$   0.18  & GALAH DR3 \\
3630462432843699200  & -1.68  $\pm$  0.10  &  0.24  $\pm$   0.05  &   0.21 $\pm$   0.10  & GALAH DR3 \\
3696393857329932672  & -1.11  $\pm$  0.38  &          --	  &   0.46 $\pm$   0.25  & GALAH DR3 \\
3753526229559762176  & -1.49  $\pm$  0.17  & -0.01  $\pm$   0.09  &  -0.00 $\pm$   0.14  & GALAH DR3 \\
6478130959611566720  & -2.38  $\pm$  0.25  &          --	  &   0.59 $\pm$   0.07  & GALAH DR3 \\
6575967187877774720  & -0.87  $\pm$  0.07  &  0.01  $\pm$   0.03  &   0.03 $\pm$   0.07  & GALAH DR3 \\
2688149803556948992  & -1.35  $\pm$  0.02  &  0.18  $\pm$   0.01  &          --      & LAMOST DR8 LR     \\
3657056526743053952  & -1.27  $\pm$  0.02  &  0.23  $\pm$   0.01  &          --      & LAMOST DR8 LR      \\
3696393857329932672  & -1.07  $\pm$  0.08  &  0.33  $\pm$   0.04  &          --      & LAMOST DR8 LR      \\
3793377208170393984  & -1.27  $\pm$  0.02  &  0.06  $\pm$   0.01  &          --      & LAMOST DR8 LR      \\
3891412241883772928  & -1.95  $\pm$  0.03  &  0.17  $\pm$   0.02  &          --      & LAMOST DR8 LR      \\
\hline
\end{tabular}
\end{table}

\section{Conclusions}\label{sect:conclusions}

In this study, we examine 64 stars that are found in both the \textit{Gaia} DR3 and the SPLUS iDR4 catalogues, with galactocentric velocities exceeding 400~$\mathrm{km\,s}^{-1}$. Given the velocities of the stars, relative to their escape velocities at their current positions, and their orbital energies, all but star \textit{Gaia} DR3 2690227738799604224 are bound to the MW. As mentioned above, due to the low SNR in radial velocity ($\mathrm{SNR}=3.23<5$) and $\varpi/\sigma_{\varpi} = 5.67 < 10$, the star \textit{Gaia} DR3 2690227738799604224 probably has an unreliable galactocentric velocity. Based on the metallicity predicted by our CNN, which is consistent with the metallicity of SPHINX, it is the most metal-poor star in our sample. Therefore, it is most likely a halo star located in the red giants branch. 

From the kinematic and dynamical analyses, all the stars show typical behavior of halo stars, and some of them have a chance of being remnants of stellar substructures, like Sequoia and GES found in the galactic halo. Confirming the association of the HiVel stars to these halo structures might provide support for the mechanism proposed by \citet{Abadi2009} to generate high velocity stars, i.e. stellar streams formed by galactic tidal forces acting on dwarf satellite galaxies during its latest pericentric passage. In addition, our analysis suggests that no star in our sample was ejected from the LMC.

From the orbital analysis, the estimated metallicity, and the location on the magnesium-metallicity diagram, the only star with a chance of originating at the GC (via Hills mechanism?) is the star \textit{Gaia} DR3 5401875170994688896 (black circle in Fig.~\ref{fig:mg_distribution}). To confirm this, an analysis of chemical abundances from high-resolution spectra is required. However, this star has been previously reported in the literature as a high-velocity star with $\mathrm{[Fe/H]} = -1.4$~dex \citep{2020A&A...638A.122C}, indicating that it is a halo star. It is worth noting, however, that the metallicity in that study was obtained from the FORS2 \citep[FOcal Reducer/low dispersion Spectrograph 2,][]{1998Msngr..94....1A} low resolution spectra ($R \sim 2\,800$), so, again, it would be interesting to carry out a study using high-resolution spectra.

While we obtain very good estimates of the effective temperatures, values of surface gravity are largely uncertain. This implies that the estimated radii should be considered with caution, and cannot be used to determine reliable values of the luminosities, comparable to the Gaia absolute magnitudes.

Finally, we would like to highlight the huge potential of applying machine learning techniques to the characterisation of stars from low resolution spectro-photometric data. In particular, we must stress the fact that this is one of the first works that is able to obtain very good [Mg/Fe] predictions for a sample of stars using such approach.

\section*{Acknowledgements}

The authors wish to thank an anonymous referee and A. Alvarez-Candal for helpful comments.

F.Q-H acknowledges financial support from CAPES. F.R., S.D., and C.B.P. acknowledge financial support from CNPq. The work of V.M.P. is supported by NOIRLab, which is managed by AURA under a cooperative agreement with the NSF. L.B.S. acknowledges the support of NASA-ATP award 80NSSC20K0509 and U.S. National Science Foundation AAG grant AST-2009122. C.M.O. thanks support from FAPESP grant 2019/26492-3.

The S-PLUS project, including the T80-South robotic telescope and the S-PLUS scientific survey, was founded as a partnership between the Fundação de Amparo à Pesquisa do Estado de São Paulo (FAPESP), the Observatório Nacional (ON), the Federal University of Sergipe (UFS), and the Federal University of Santa Catarina (UFSC), with important financial and practical contributions from other collaborating institutes in Brazil, Chile (Universidad de La Serena), and Spain (Centro de Estudios de Física del Cosmos de Aragón, CEFCA). We further acknowledge financial support from the São Paulo Research Foundation (FAPESP), Fundação de Amparo à Pesquisa do Estado do RS (FAPERGS), the Brazilian National Research Council (CNPq), the Coordination for the Improvement of Higher Education Personnel (CAPES), the Carlos Chagas Filho Rio de Janeiro State Research Foundation (FAPERJ), and the Brazilian Innovation Agency (FINEP). The authors who are members of the S-PLUS collaboration are grateful for the contributions from CTIO staff in helping in the construction, commissioning and maintenance of the T80-South telescope and camera. We are also indebted to Rene Laporte and INPE, as well as Keith Taylor, for their important contributions to the project. From CEFCA, we particularly would like to thank Antonio Marín-Franch for his invaluable contributions in the early phases of the project, David Cristóbal-Hornillos and his team for their help with the installation of the data reduction package jype version 0.9.9, César Íñiguez for providing 2D measurements of the filter transmissions, and all other staff members for their support with various aspects of the project.

This work used of data from the European Space Agency (ESA) mission {\it Gaia} (\url{https://www.cosmos.esa.int/gaia}), processed by the {\it Gaia} Data Processing and Analysis Consortium (DPAC, \url{https://www.cosmos.esa.int/web/gaia/dpac/consortium}). Funding for the DPAC has been provided by national institutions, in particular the institutions participating in the {\it Gaia} Multilateral Agreement.

\section*{Data Availability}
 
The data related to the results presented in this work and the trained CNN and ANN can be available on reasonable request to the corresponding author.


\bibliographystyle{mnras}
\bibliography{splus_hivel} 


\appendix
\section{Distances}\label{apend:distances}

Because $V_{\mathrm{GC}}$ is sensitive to the stellar distances, we compare the $V_{\mathrm{GC}}$ computed using the photo-astrometric distances of \citet{2022A&A...658A..91A} and the photo-geometric distances of \citet{2021yCat.1352....0B}. Both distances were obtained applying a probabilistic approach. Photo-geometric distances were estimated using parallaxes, \textit{Gaia} photometry, and a prior information constructed from a three-dimensional model of the MW. Photo-astrometric distances were estimated using parallaxes, \textit{Gaia} photometry, information from photometric catalogues, and the stellar-density prior of the StarHorse code. Section 3 of \citet{2022A&A...658A..91A} discusses the comparison between the distances for 1 million random stars, and they show a good agreement. 

For our sample, the distances show a good agreement (mean offset null) for distances $\lesssim 7$~kpc. Beyond this value, the StarHorse distances for five stars appear underestimated (mean of $\sim 2.6$~kpc) relative to the Bailer-Jones distances. Fig.~\ref{fig:escape-vel-curve-sh-bj} shows the $V_{\mathrm{GC}}$ as a function of $R_{\mathrm{GC}}$ for the HiVel sample, using the photo-geometric distances (red circles) and the photo-astrometric distances (blue circles). We see that the star \textit{Gaia} DR3 3631125674577631104 is an unbound candidate if we consider the StarHorse distance, but with a large uncertainty in $V_{\mathrm{GC}}$.
   
\begin{figure}
    \centering
    \includegraphics[scale=.7]{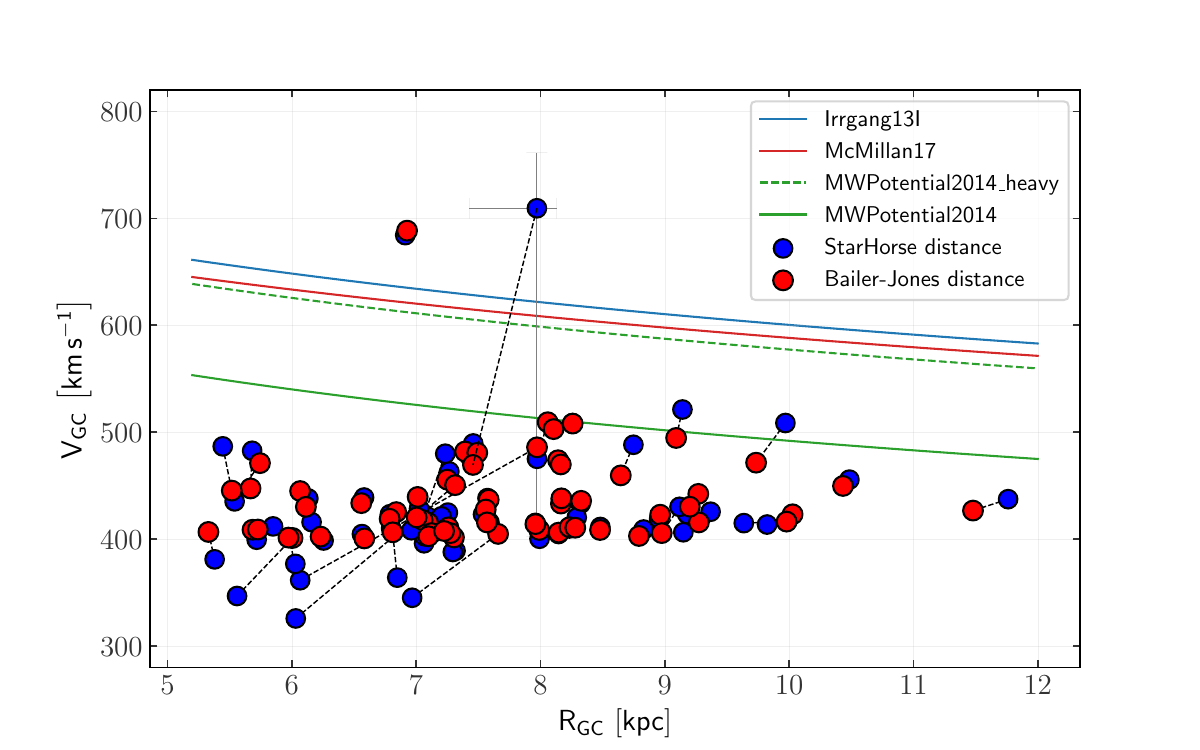}
    \caption{Galactocentric velocity as a function of Galactocentric distance. Red circles computed with the photo-geometric distances \citep{2021yCat.1352....0B}, and blue circles computed with the photo-astrometric distances \citep{2022A&A...658A..91A}.}
    \label{fig:escape-vel-curve-sh-bj}
\end{figure}

\bsp	
\label{lastpage}
\end{document}